\documentclass[11pt, a4paper]{article}
\pdfoutput=1

\usepackage{amsmath}
\usepackage{amsfonts}
\usepackage{amssymb}
\usepackage{bm}
\usepackage{bbm}
\usepackage{verbatim}
\usepackage{dsfont}
\usepackage{booktabs}
\usepackage{slashed}
\usepackage{nicefrac}
\usepackage{mathtools}
\usepackage{physics}
\usepackage{array}

\usepackage{epsfig}
\usepackage[dvipsnames]{xcolor}
\usepackage{graphicx}
\usepackage[font=small]{caption}
\usepackage[listofformat=empty,subrefformat=empty]{subfig}

\usepackage{float}
\usepackage{overpic}
\usepackage{tikz}
\usepackage{tabularx}

\usepackage{enumerate}
\usepackage{hhline}
\usepackage{multirow}

\usepackage[nosort]{cite}
\usepackage{xspace}
\usepackage{setspace}

\usepackage[pdftitle={},
  pdfauthor={},
  pdfsubject={},
  bookmarksopen, bookmarksnumbered, bookmarksopenlevel=2, colorlinks=false, linkcolor=blue, citecolor=blue, urlcolor=blue]{hyperref}

\usepackage[nameinlink,capitalize]{cleveref}

\makeatletter{}
\g@addto@macro\bfseries{\boldmath}
\makeatother

\newcommand*{\bbZ}{\ensuremath{\mathbb{Z}}}

\usepackage[normalem]{ulem}

\newcommand{\be}{\begin{equation}}
\newcommand{\ee}{\end{equation}}
\newcommand{\ba}{\begin{aligned}}
\newcommand{\ea}{\end{aligned}}

\def\half{{\frac{1}{2}}}

\def\Z{{\mathbb{Z}}}
\def\R{{\mathbb{R}}}

\newcommand{\rr}{\rightarrow}

\begin{document}

\baselineskip=18pt  
\numberwithin{equation}{section}  
\allowdisplaybreaks  


\thispagestyle{empty}

\vspace*{-2cm}

\vspace*{2.6cm}
\begin{center}
{\Large{\textbf{Decomposition and (Non-Invertible) $(-1)$-Form Symmetries \vspace{.4cm} \\
from the Symmetry Topological Field Theory \vspace{.4cm}
}}} \\
\vspace*{1.5cm}
Ling Lin$^{1,2}$, Daniel Robbins$^3$, Subham Roy$^3$\\

\vspace*{1.0cm}
{\it $^1$Dipartimento di Fisica e Astronomia, Universit\`{a} di Bologna, via Irnerio 46, Bologna, Italy}\\
\vspace*{.1cm}
{\it $^2$INFN, Sezione di Bologna, viale Berti Pichat 6/2, Bologna, Italy}\\
\vspace*{.3cm}
{\it $^3$Department of Physics, University at Albany, Albany, USA}

\vspace*{0.8cm}
\end{center}
\vspace*{.5cm}

\noindent
We investigate the interplay between $(-1)$-form symmetries and their quantum-dual $(d-1)$-form counterparts within the framework of Symmetry Topological Field Theories (SymTFTs).
In this framework the phenomenon of decomposition --- a $d$-dimensional quantum field theory with $(d-1)$-form symmetry being the disjoint union of other theories (or ``universes'') --- arises naturally from manipulations of topological boundary conditions of the SymTFT.
We corroborate our findings with various examples, including a generalization of ``instanton-restricted'' 4d Yang--Mills theories with no sum over instanton sectors.
Furthermore, we construct a 3d SymTFT with a non-invertible $(-1)$-form symmetry.
The absolute 2d quantum field theory includes a 0-form global symmetry that depends on a parameter whose value gets shifted by the action of the $(-1)$-form symmetry, and we show that the non-invertibility of the latter is needed to encode this modification of the 0-form symmetry.

\newpage
\setcounter{tocdepth}{2}
\tableofcontents

\newpage


\section{Introduction}

The Symmetry Topological Field Theory (SymTFT) \cite{Freed:2012bs,Gaiotto:2020iye,Apruzzi:2021nmk,Freed:2022qnc} (see also \cite{Schafer-Nameki:2023jdn, Bhardwaj:2023kri} for reviews and further references) is a unified framework to study generalized global symmetries of quantum field theories (QFTs) \cite{Gaiotto:2014kfa}.
The underlying principle is that topological symmetry generators and extended charged objects of a $d$-dimensional QFT originate democratically from a $(d+1)$-dimensional topological bulk.
This framework is particularly powerful in encoding different ``global forms'' of a QFT:
the emergence of a quantum-dual symmetry $\widehat{G}$ (with its anomalies and any other categorical bells and whistles) from the gauging of a global symmetry $G$ is succinctly captured by change of topological boundary conditions for the $(d+1)$-dimensional SymTFT.

In this work we investigate $(-1)$-form symmetries and their quantum-dual $(d-1)$-form symmetries through the SymTFT-lens.
These ``extremal'' cases of higher-form symmetries have been previously studied mostly separately.
While $(-1)$-form symmetries have been discussed as a symmetry on ``coupling parameters'' \cite{Cordova:2019jnf, Cordova:2019uob, Heidenreich:2020pkc, Vandermeulen:2022edk, Aloni:2024jpb} , $(d-1)$-form symmetries have appeared in the context of orbifolds and the phenomenon of decomposition \cite{Hellerman:2006zs, Sharpe:2014tca, Sharpe:2019ddn, Tanizaki:2019rbk, Komargodski:2020mxz, Cherman:2020cvw, Robbins:2020msp, Nguyen:2021naa, Sharpe:2021srf, Cherman:2021nox, Sharpe:2022ene,Lin:2022xod, Bhardwaj:2023bbf}.
Our primary goal is to connect these perspectives in a natural way through the SymTFT for both finite and infinite Abelian symmetries.
Moreover, we also use the SymTFT to construct an example of non-invertible $(-1)$-form symmetry, which to our knowledge has not been studied before.

\subsubsection*{(Non-invertible) $(-1)$-form symmetries}

From the perspective of a $d$-dimensional QFT, the usual definition of a $p$-form symmetry becomes somewhat tenuous for $p=-1$.
On one hand, $(-1)$-dimensional charged objects arguable cannot exist, and on the other hand any spacetime filling operator appears to be trivially topological.
Instead these subtleties, including the distinction between $\mathbb{Z}_N$ and $U(1)$ $(-1)$-form symmetries, are all elegantly encoded from a $(d+1)$-dimensional point of view through the SymTFT of the $d$-dimensional theory.

Generically, this $(d+1)$-dimensional theory contains topological point operators $U(x)$ and codimension-1 (domain wall) operators $V(\Gamma_d)$ which are canonically conjugate in the Hamiltonian picture, resulting in a complex phase as one drags the point operator across the domain wall $V(\Gamma_d)$.
For such a system, we have again the standard gapped boundary conditions with Dirichlet / Neumann for either $U$ / $V$ or for $V$ / $U$.

Dirichlet boundary conditions for the point operators simply means that $U(x)$ takes a fixed value $\hat{U}(x)$; in typical examples, where $U(x) = \exp(i b_0(x))$ for a scalar field $b_0(x)$ in the bulk, the fixed value $b_0(x)|_{\mathfrak{B}} \equiv \hat{b}$ is identified with the value of a (coupling) ``parameter'' of the $d$-dimensional QFT.
The resulting global form of the $d$-dimensional theory has a $(-1)$-form symmetry: the insertion of the now spacetime-filling operators $V$ which have Neumann boundary conditions ``acts'' by shifting the boundary value of $U$ (i.e., the parameter $\hat{b}$).

This standard SymTFT perspective provides an interesting re-interpretation of the symmetry action that is unique to $(-1)$-form symmetries: shifting the coupling parameters is achieved by stacking the $d$-dimensional theory with a topological sector, which in the SymTFT description corresponding to worldvolume of the $V$ operator.
By definition, this sector is trivial precisely when the corresponding symmetry is invertible.
Many previously studied examples, including recent top-down string models \cite{Yu:2024jtk, Najjar:2024vmm, Najjar:2025htp}, fall into this category, where the $(-1)$-form symmetry merely shifts a ``parameter''.

One of the goals of this work is to present a non-invertible generalization in $d=2$.
In this example, changing the ``parameter'' on which the $(-1)$-form symmetry acts also modifies a 0-form global symmetry of the 2d theory.
As we will see in detail, the non-invertibility of the 2d defects $V$ is necessary to incorporate both the $(-1)$-form symmetry and the varying 0-form symmetry within a \emph{single} SymTFT.

\subsubsection*{Decomposition from the SymTFT}

If we instead impose Dirichlet boundary conditions for the codimension-1 operators, we obtain the global form with a $(d-1)$-form symmetry that is generated by the topological point operators $U$ which have Neumann boundary conditions.
The existence of the point operators is linked to ``decomposition'' which we now briefly review (see also \cite{Sharpe:2022ene}).

A $d$-dimensional theory $\widetilde{\cal T}$ is said to decompose if there are topological operators $\{\Pi_i (x)\}$ which satisfy $\Pi^2 = \Pi$, i.e., they act as projections onto disjoint sectors of the theory when inserted into correlation functions (such as the partition function $Z$):
\begin{align}\label{eq:decomp_partition-function}
    \begin{split}
        \widetilde{\cal T} = \bigoplus_i {\cal T}_i \, , \quad \widetilde{\cal H} = \bigoplus_i {\cal H}_i \, , \quad \widetilde{Z} = \sum_i Z_i \, .
    \end{split}
\end{align}
The summands are themselves $d$-dimensional QFTs ${\cal T}_i$ with partition function $Z_i$, which are completely disjoint from each other: there are no finite energy processes (even in a spacetime of finite volume) which maps between their Hilbert spaces ${\cal H}_i$.
Each summand is called a ``universe''.
In the modern language of generalized symmetries, the projectors are certain linear combinations of the topological point operators $U$ that generate the $(d-1)$-form symmetry  of $\widetilde{\cal T}$.
In the simplest instance where this is a group-like symmetry $G^{(d-1)}$, the projection onto the individual ``universes'' ${\cal T}_i$ can be equivalently thought of as a (topological) gauging of $G^{(d-1)}$, with ${\cal T}_i$'s corresponding to different choices of discrete torsion parameters $\hat{b}_i \in H^{d-(d-1)}(G^{(d-1)}, U(1))$ for this gauging.
In general, the (discrete) gauging of a group-like $p$-form symmetry $G^{(p)}$ results in a quantum-dual $(d-p-2)$-form symmetry $\widehat{G}^{(d-p-2)}$, where $\widehat{G}$ is the Pontryagin-dual of $G$.
The universes appearing in decomposition (associated to a group-like $(d-1)$-form symmetry) \eqref{eq:decomp_partition-function} will thus have a $(-1)$-form symmetry $\widehat{G}^{(-1)}$.

As we can anticipate, this comes out naturally from the SymTFT through a change of boundary conditions.
In particular, the discrete torsion parameter $\hat{b}_i$ that label the universes are identified with the Dirichlet boundary values of the scalar field $b_0$ which becomes the ``parameter'' on which the $(-1)$-form symmetry acts on.
Moreover, we shall demonstrate that this change, which can be equivalently viewed as a basis transformation in the Hilbert space of the SymTFT, reproduces on the nose the defining properties \eqref{eq:decomp_partition-function} of decomposition.
Importantly, with the developments in \cite{Antinucci:2024zjp, Brennan:2024fgj, Bonetti:2024cjk, Apruzzi:2024htg} we can also describe a decomposition into \emph{infinitely many} universes associated with a $\widehat{U(1)} = \mathbb{Z}$ $(d-1)$-form symmetry.
One application that we will highlight is a generalization of the ``instanton restriction'' of 4d Yang--Mills theories \cite{Tanizaki:2019rbk} to a single instanton sector, which decomposes into a continuum of ``universes'' with parameter being the theta angle $\theta \in U(1)$.

Another way that decomposition arises is from the gauging of a trivially acting $(d-2)$-form symmetry \cite{Pantev:2005rh, Robbins:2020msp, Robbins:2021lry, Robbins:2021ibx,Pantev:2022kpl,Pantev:2022pbf, Robbins:2022wlr}.  The triviality of the symmetry means that the associated topological line operators cannot link with anything else and there are no operators charged under this symmetry.  This is accomplished by ensuring that the trivial line operators can end on topological point operators --- in this way the line operator can always be ``unhooked'' from any linking.  Upon gauging the line operators become transparent (in the absolute theory), but the point operators can persist, becoming the point operators $U(x)$ of decomposition.  Again, this has an elegant description in the SymTFT language.

\subsubsection*{Organization of the paper}

The paper is organized as follows.
We begin in Section \ref{sec:BF-SymTFT} with a general discussion of group-like $(-1)$- and $(d-1)$-form symmetries, including infinite groups $U(1)^{(-1)}$ and $\mathbb{Z}^{(d-1)}$, within the SymTFT and derive the hallmarks \eqref{eq:decomp_partition-function} of decomposition from it.
Section \ref{sec:instanton_4d} presents a concrete example in form of the instanton-restricted 4d Yang--Mills theory where we highlight decomposition with a $\mathbb{Z}^{(3)}$ symmetry.
In Section \ref{sec:trivially_acting} we study SymTFTs for trivially acting $(d-2)$-form symmetries in conrete 2d examples, where we provide an alternative to the ``club sandwich'' description \cite{Bhardwaj:2023bbf,Bhardwaj:2024qrf}.
We then utilize the full power of the SymTFT in Section \ref{sec:non-invertible} to construct a 2d model with a non-invertible $(-1)$-form symmetry, and investigate in detail how it modifies the 0-form symmetry captured within the same bulk theory.
We conclude in Section \ref{conclusion} with possible future directions that would arise naturally from this work.

\section{BF-SymTFT for \texorpdfstring{\boldmath{$(-1)$}}{(-1)}-form symmetries and decomposition}\label{sec:BF-SymTFT}

We start with a brief review of the SymTFT framework (see \cite{Schafer-Nameki:2023jdn, Bhardwaj:2023kri} for more comprehensive ones).
In a nutshell, the SymTFT $\mathfrak{Z}$ associated to a $d$-dimensional quantum field theory $\widetilde{\cal T}$ on ${\cal M}_d$ is a topological $(d+1)$-dimensional theory on ${\cal W}_{d+1} \cong {\cal M}_d \times [0,1]$.
We will assume throughout the paper that ${\cal M}_d$ is connected.
The boundary at $r=1$, where $r$ is the interval coordinate, is called the physical boundary $\mathfrak{B}_\text{phys} \cong {\cal M}_d$, and contains the gapless degrees of freedom of $\widetilde{\cal T}$.
The global symmetries and generalized charges are encoded in the choice of gapped boundary conditions for the topological operators of $\mathfrak{Z}$ at $r=0$, referred to as the symmetry boundary $\mathfrak{B}_\text{sym} \cong {\cal M}_d$.\footnote{We will use $\mathfrak{B}_\text{sym}$ to denote both the boundary of ${\cal W}_{d+1}$ that is homeomorphic to the $d$-dimensional spacetime as well as the boundary condition for the operators.}
$\widetilde{\cal T}$ is obtained by reducing the interval, or collapsing the ``sandwich''.
Other ``global variants'' ${\cal T}_i$ of the $d$-dimensional theory, related to $\widetilde{\cal T}$ by gauging some of its global symmetries labeled by some discrete torsion parameters (with label $i$) are obtained by different choices of symmetry boundary conditions $\mathfrak{B}_\text{sym}$.

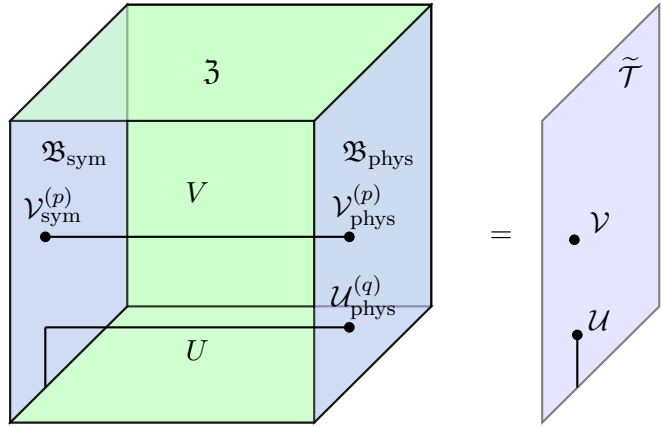
\begin{figure}[ht]
  \centering
    \begin{tikzpicture}
    \coordinate (A) at (0,0,0);
    \coordinate (B) at (4,0,0);
    \coordinate (C) at (4,4,0);
    \coordinate (D) at (0,4,0);
    \coordinate (E) at (0,0,4);
    \coordinate (F) at (4,0,4);
    \coordinate (G) at (4,4,4);
    \coordinate (H) at (0,4,4);

    \draw[thick, fill=green!20] (A) -- (B) -- (C) -- (D) -- cycle; 
    \draw[thick, fill=green!20, opacity=0.7] (E) -- (F) -- (G) -- (H) -- cycle; 
    \draw[thick, fill=green!20, opacity=0.7] (A) -- (B) -- (F) -- (E) -- cycle; 
    \draw[thick, fill=blue!20, opacity=0.7] (B) -- (C) -- (G) -- (F) -- cycle; 
    \draw[thick, fill=blue!20, opacity=0.7] (A) -- (D) -- (H) -- (E) -- cycle; 
    \draw[thick, fill=green!20, opacity=0.7] (D) -- (C) -- (G) -- (H) -- cycle; 

    \node at (1.5, 3.5, 1.0) {$\mathfrak{Z}$}; 
    \node at (-0.5, 2.2, 0.5) {$\mathfrak{B}_\text{sym}$}; 
    \node at (3.5, 2.2, 0.5) {$\mathfrak{B}_\text{phys}$}; 

    \draw[thick] (-0.5, 1.5, 1.5) -- (3.5, 1.5, 1.5);
    \node at (1.5, 2.1, 1.5) {$V$};

    \node at (-0.4, 1.9, 1.5) {$\mathcal{V}_\text{sym}^{(p)}$};
    \node at (3.7, 1.9, 1.5) {$\mathcal{V}^{(p)}_\text{phys}$};

    \fill[black] (-0.5, 1.5, 1.5) circle (2pt);
    \fill[black] (3.5, 1.5, 1.5) circle (2pt);

    \draw[thick] (-0.5, 0.3, 1.5) -- (3.5, 0.3, 1.5);
    \node at (1.5, 0.0, 1.5) {$U$};
    \draw[thick] (-0.5, 0.3, 1.5) -- (-0.5, -.5, 1.5);

    \fill[black] (3.5, 0.3, 1.5) circle (2pt);
    \node at (3.7, 0.7, 1.5) {$\mathcal{U}_\text{phys}^{(q)}$};
    
        \node at (5.5, 1.5, 1.5) {$=$};

        \begin{scope}[shift={(7,0,0)}] 
            \coordinate (P1) at (0,0,0);
            \coordinate (P2) at (0,0,4);
            \coordinate (P3) at (0,4,4);
            \coordinate (P4) at (0,4,0);
            \draw[thick, fill=blue!20, opacity=0.5] (P1) -- (P2) -- (P3) -- (P4) -- cycle; 

            \fill[black] (0, 2, 2.9) circle (2pt);
            \node at (0.3, 2.2, 2.8) {${\cal V}$}; 
            \node at (0.0, 3.5, 1.0) {$\widetilde{\cal T}$};       

            \fill[black] (0, 0.7, 2.8) circle (2pt);
            \draw[thick] (0, 0.7, 2.8) -- (0, 0.0, 2.8);
            \node at (0.3, 0.9, 2.8) {${\cal U}$};
        \end{scope}
\end{tikzpicture}
\caption{A cartoon depiction of the SymTFT construction.
The $V$ operator is shown with Dirichlet boundary conditions on $\mathfrak{B}_\text{sym}$.
The end-``points'' on both boundaries together give rise to a gauge-invariant genuine operator ${\cal V}$ charged under a $p$-form symmetry after closing the interval.
On the other hand, the Neumann boundary condition for $U$ forces it to continue on $\mathfrak{B}_\text{sym}$, resulting in a non-genuine $q = (d-p-2)$-form charge ${\cal U}$.
}
\label{fig:SymTFT}
\end{figure}

To describe invertible Abelian $p$-form symmetries $G^{(p)}$ and their quantum dual $(d-p-2)$-form symmetries $\widehat{G}^{(d-p-2)}$, the $(d+1)$-dimensional bulk theory contains topological operators $U^{(d-p-1)} \equiv U(\Sigma_{d-p-1})$ and $V^{(p+1)} \equiv V(\Gamma_{p+1})$ of the indicated dimension.
Their linking phase,
\begin{align}\label{eq:commutation_BF-operators}
    U(\Sigma_{d-p-1}) V(\Gamma_{p+1}) = \exp(2\pi i \, \alpha \, \text{Link}(\Sigma, \Gamma)) V(\Gamma_{p+1}) U(\Sigma_{d-p-1}) \, ,
\end{align}
can be derived from a BF-type action in the bulk, see Appendix \ref{app:symTFT_reviews}. 
$V^{(p+1)}$ and $U^{(d-p-1)}$ can end ``perpendicularly'' on $\mathfrak{B}_\text{phys}$, which give rise to the (generally non-mutually local) charged objects ${\cal V}^{(p)}$ and ${\cal U}^{(d-p-2)}$ of $G^{(p)}$ and $\widehat{G}^{(d-p-2)}$, respectively.
A consistent choice of boundary conditions for the operators $U$ and $V$ on $\mathfrak{B}_\text{sym}$ selects a maximal mutually local set of operators (ending with Dirichlet boundary conditions) that survive the interval reduction as genuine charged objects ${\cal U}$ and ${\cal V}$; meanwhile the Pontryagin-dual set (with Neumann boundary conditions) can lie parallel to the boundary and generate the global symmetries of the $d$-dimensional theory after interval reduction.
Figure \ref{fig:SymTFT} provides a pictorial summary.

To perform a symmetry transformation in the $d$-dimensional theory $\widetilde{\cal T}$, say of its $G^{(p)}$ symmetry, one inserts the $U$ operator, which has Neumann boundary conditions on $\mathfrak{B}_\text{sym}$, along a cycle $\Sigma_{d-p-1} \subset \mathfrak{B}_\text{sym}$, which may link with $V$ insertions on $\Gamma_{p+1}^{(i)}$ that have to end perpendicularly on $\gamma_{p}^{(i)} \subset \mathfrak{B}_\text{sym}$, i.e., insertions of operators that are charged under the $p$-form symmetry in the $d$-dimensional theory.
At the level of correlation functions we have
\begin{align}\label{eq:symmetry_transformation_insertions}
    \begin{split}
        \langle {\textstyle\prod_i} {\cal V} (\gamma_p^{(i)})' \, \cdots \, \rangle_{\widetilde{\cal T}} = \langle U(\Sigma_{d-p-1}) \, {\textstyle\prod_i} {\cal V} (\gamma_p^{(i)}) \, \cdots \, \rangle_{\widetilde{\cal T}} = \langle U(\Sigma_{d-p-1}) \, {\textstyle\prod_i} V (\Gamma_{p+1}^{(i)}) \, \cdots \, \rangle_\mathfrak{Z} \, ,
    \end{split}
\end{align}
where `$\cdots$' denote possible further insertions of ``uncharged'' operators, including those localized in the gapless sector on $\mathfrak{B}_\text{phys}$, and where a prime indicates a symmetry transformed charged operator.
By the topological nature of $\mathfrak{Z}$, this is equivalent to inserting $U(\Sigma_{d-p-1})$ at any value of the interval coordinate $r$, i.e., anywhere in the interior of the bulk ${\cal W}_{d+1} \cong {\cal M}_d \times [0,1]$.

Given our motivations to study $(-1)$-form symmetries and decomposition, we will focus now on the case $p=d-1$ (or, equivalently, $p=-1$, and exchange the $U$ and $V$ operators).
In this case, $U^{(d-p-1)} \equiv U(x)$ are topological point operators, and $V^{(p+1)} \equiv V(\Gamma_d)$ are topological codimension-1, or domain walls, in the $(d+1)$-dimensional SymTFT.

Almost all of the features of the general discussion above apply in this case.
In particular, non-trivial linking between $U(x)$ and $V(\Gamma_d)$ leads to the usual constraints on boundary conditions (see Appendix \ref{app:symTFT_reviews} for a review of the details).
The only difference is that the ``geometric'' intuition of Dirichlet boundary conditions as ``ending a defect perpendicularly'' on the boundary is not applicable to the point operators $U(x)$; that is, there are no ``charged operators'' ${\cal U}^{(-1)}$ for $(-1)$-form symmetries in the $d$-dimensional theory.
Instead, we will see that it is more convenient to simply think of Dirichlet boundary conditions, i.e., a fixed value for $U(x)$ on the boundary, as a ``label'' for different \emph{universes}.

The action of the $(-1)$-form symmetry is analogous to the general case described above.
In the present case, the symmetry generators $V$, being a codimension 1 defect \emph{in the bulk}, actually fills the entire boundary spacetime, consistent with the expectation that a $(-1)$-form symmetry in $d$ dimensions is generated by insertions codimension-0 topological defects.
In accordance with the linking property, such an insertion of $V$ shifts the Dirichlet-fixed value of $U(x)$, thereby leading to an absolute theory corresponding to a different universe.\footnote{This is the most ``traditional'' version of $(-1)$ form symmetry. However, we believe there may be more exotic versions, where the codimension-one operator in the bulk does not act on any point operator but may still exhibit higher charges.
We leave the discussion of such versions for future work.}

\subsection{Universes and spacetime-filling defects with \texorpdfstring{$\mathbb{Z}_N$}{Z/N}-SymTFT}

The simplest example of a $(d+1)$-dimensional SymTFT with topological points and domain walls is given by a $G \cong \widehat{G} = \mathbb{Z}_N$ BF-theory with a 0-form and a $d$-form gauge field,
\begin{align}\label{eq:BF-with-0form}
    S_\text{bulk}^{(d+1)} \equiv S_\text{bulk} = \frac{i N}{2\pi} \int_{{\cal W}_{d+1}} b_0 \wedge dc_d = \frac{-iN}{2\pi} \int_{{\cal W}_{d+1}} db_0 \wedge c_d + \frac{iN}{2\pi} \int_{\partial {\cal W}_{d+1}} b_0 \wedge c_d \, .
\end{align}
From the equations of motion we have\footnote{Here $b_0$ and $c_d$ are $U(1)$-valued fields, meaning in particular that $b_0$ is a $2\pi$-periodic scalar even before imposing equations of motion, so $n$ is only sensibly defined mod $N$.} $N \oint c_d = N b_0 = 0 \mod 2 \pi$, i.e., $b_0(x) = 2\pi n/N$, $n=0,..., N-1$, is locally constant.
Therefore, the point operators $U(x)=\exp(i b_0(x))$ are topological.
The codimension-1 operators $V(\Gamma_d;k) = \exp(i k \oint_{\Gamma_d} c_d)$ then behave as (topological) domain walls, across which the value of $b_0$ shifts by $2\pi \tfrac{k}{N}$.
To see this, imagine $\Gamma_d$ being located at $x_r = 0$ for the $r$-th coordinate, and we have chosen an orientation in $(d+1)$-dimensions such that a point ``to the left'' of $\Gamma_d$ ($x_r<0$) links positively with $\Gamma_d$, i.e., $\text{Link}(\{x \, | \, x_r < 0\}, \Gamma_d) =1$.
Then the standard commutation relation \eqref{eq:commutation_BF-operators} between the $U$ and $V$ operators becomes
\begin{align}\label{eq:TPOs_jump_domain-wall}
\begin{split}
    & \exp(i \, b_0(x_r = +\epsilon)) \, V(\Gamma_d;k) = \exp(2\pi i \, \tfrac{k}{N} ) V(\Gamma_d;k) \exp(i\,b_0(x_r=-\epsilon)) \\ 
    \Rightarrow \quad & b_0(x_r= +\epsilon) = b_0(x_r = -\epsilon) + 2\pi \tfrac{k}{N} \, .
\end{split}
\end{align}
Note that we can equally interpret this equation as the action of the $U$-operator on $V(\Gamma_d)$, which shifts the periods of $c_d$,
\begin{align}\label{eq:U_shift_c-periods}
    U \, V(\Gamma_d;1) \, U^{-1} = U \, \exp(i \textstyle\int_{\Gamma_d} c_d) \, U^{-1} = \exp(i (\textstyle\int_{\Gamma_d} c_d + \tfrac{2\pi}{N} )) \, .
\end{align}
Next we will describe the standard boundary conditions for this system that correspond to a $d$-dimensional theory with either a $(d-1)$-form or a $(-1)$-form symmetry.

\subsubsection*{Boundary conditions}

The theory with $(d-1)$-form symmetry arises from Dirichlet boundary conditions for $V(\Gamma_d;k)$, resulting in codimension-1 defects ${\cal V}(\partial \Gamma_d;k)$ in the $d$-dimensional QFT, with $\partial \Gamma_d$ contained in the boundary.
These are the charged objects under the $(d-1)$-form symmetry generated by $U(x;k) = \exp(i \, k \, b_0(x))$, with $x$ a point on the boundary.
Notice that the ``symmetry transformation'' on ${\cal V}(\partial \Gamma_d;k)$ inherited from \eqref{eq:TPOs_jump_domain-wall} can be again interpreted as ${\cal V}$ being domain walls for the field $b_0$ in $d$ dimensions.
One way to think about the different universes from decomposition associated with the $(d-1)$-form symmetry are then the regions separated by these domain walls.
The locally constant field $b_0$ with Neumann boundary condition will then jump across the domain walls.

\begin{figure}[ht]
  \centering
    \begin{tikzpicture}
    \coordinate (A) at (0,0,0);
    \coordinate (B) at (4,0,0);
    \coordinate (C) at (4,4,0);
    \coordinate (D) at (0,4,0);
    \coordinate (E) at (0,0,4);
    \coordinate (F) at (4,0,4);
    \coordinate (G) at (4,4,4);
    \coordinate (H) at (0,4,4);

    \draw[thick, fill=green!20] (A) -- (B) -- (C) -- (D) -- cycle; 
    \draw[thick, fill=green!20, opacity=0.7] (E) -- (F) -- (G) -- (H) -- cycle; 
    \draw[thick, fill=green!20, opacity=0.7] (A) -- (B) -- (F) -- (E) -- cycle; 
    \draw[thick, fill=blue!20, opacity=0.7] (B) -- (C) -- (G) -- (F) -- cycle; 
    \draw[thick, fill=blue!20, opacity=0.7] (A) -- (D) -- (H) -- (E) -- cycle; 
    \draw[thick, fill=green!20, opacity=0.7] (D) -- (C) -- (G) -- (H) -- cycle; 

    \node at (1.5, 3.5, 1.0) {$\mathfrak{Z}$}; 
    \node at (-0.5, 2.2, 0.5) {$\mathfrak{B}_\text{sym}$}; 
    \node at (3.5, 2.2, 0.5) {$\mathfrak{B}_\text{phys}$}; 

    \draw[fill=red!20, opacity = .5] (0, 1.5, 0) -- (4, 1.5, 0) -- (4, 1.5, 4) -- (0, 1.5, 4) -- cycle;

    \node at (1.5, 1.5, 1.5) {$V(\Gamma_d;k)$};

    \node at (-0.4, 1.9, 1.5) {$\mathcal{V}_\text{sym}$};
    \node at (3.7, 1.9, 1.5) {$\mathcal{V}_\text{phys}$};

    \draw[thick] (0, 1.5, 0) -- (0, 1.5, 4);
    \draw[thick] (4, 1.5, 0) -- (4, 1.5, 4);

    \node at (5.5, 1.5, 1.5) {$=$};

    \begin{scope}[shift={(7,0,0)}] 
        \coordinate (P1) at (0,0,0);
        \coordinate (P2) at (0,0,4);
        \coordinate (P3) at (0,4,4);
        \coordinate (P4) at (0,4,0);
        \draw[thick, fill=blue!20, opacity=0.5] (P1) -- (P2) -- (P3) -- (P4) -- cycle; 

        \draw[thick] (0, 1.5, 0) -- (0, 1.5, 4);
        \node at (0.3, 2.2, 2.8) {${\cal V}$}; 
        
        \node at (0.8, 0.8, 2.8) {$b_0 = 2\pi \tfrac{n}{N}$};
        \node at (0.8, 3.2, 2.8) {$b_0 = 2\pi \tfrac{n+k}{N}$};
    \end{scope}
\end{tikzpicture}
\caption{Dirichlet boundary conditions for $V$ allows for operator insertions which leads to a codimension-1 domain wall in the $d$-dimensional theory, across which the locally constant scalar $b_0$ jumps.}
\label{fig:dirichlet_for_V}
\end{figure}
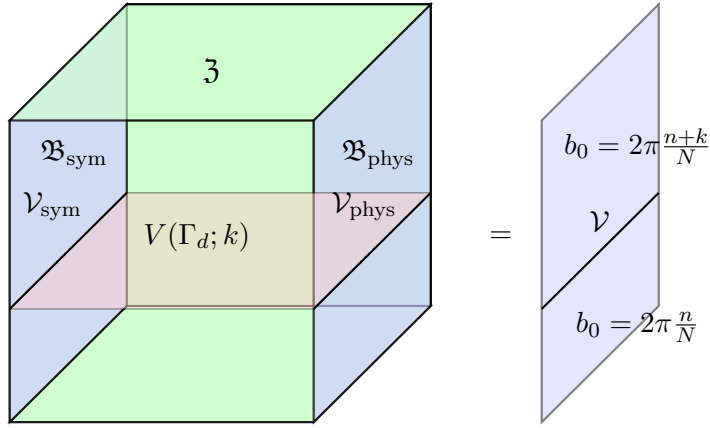

As reviewed in Appendix \ref{app:symTFT_reviews}, this boundary condition corresponds to a state $|\mathfrak{B}^{(\hat{c})} \rangle$ of the $(d+1)$-dimensional SymTFT, which can be described by a topological action on the $d$-dimensional symmetry boundary,
\begin{align}\label{eq:sym-action_for_decomp}
    |\mathfrak{B}^{(\hat{c})} \rangle : \quad S_\text{sym}[\lambda_{d-1}] = \frac{-iN}{2\pi} \int_{\mathfrak{B}_\text{sym}} b_0 \wedge (\hat{c}_d - d \lambda_{d-1})\, .
\end{align}
Here, $\lambda_{d-1}$ is a dynamical field (i.e., path-integrated over) that is confined to $\mathfrak{B}_\text{sym}$, while $b_0 \equiv b_0 |_{\mathfrak{B}_\text{sym}}$ is the restriction of the scalar field.
Meanwhile, $\hat{c}_d$ is a \emph{fixed} gauge field that becomes the Dirichlet-boundary value of the bulk $d$-form field up to gauge transformations,
\begin{align}
    c_d |_{\mathfrak{B}_\text{sym}} = \hat{c}_d - d\lambda_{d-1} \, .
\end{align}
This is simply the equations of motion for the bulk-boundary system with bulk action \eqref{eq:BF-with-0form}.
This leaves $b_0$ as dynamical field which is summed over in the path-integral of the $d$-dimensional QFT after interval reduction.
Because $b_0$ is locally constant, the path-integral, in the absence of the domain walls, literally decomposes into a sum over different possible values $b_0$ can take within a connected component of spacetime, i.e., universes.

Notice that, after colliding with the physical boundary in the process of the interval reduction, we end up with a $d$-dimensional theory which will generally contain \eqref{eq:sym-action_for_decomp} as part of its action,
\begin{align}\label{eq:general_decomposing_lagrangian}
    S^{(d)}[..., b_0, \lambda_{d-1}] \supset \frac{iN}{2\pi} \int_{{\cal M}_d}  b_0 \wedge ( \hat{c}_d - d\lambda_{d-1} ) \, ,
\end{align}
where we recognize a BF-pair $(b_0, \lambda_{d-1})$.
In $d$-dimensions, the $(d-1)$-form symmetry is thus realized as holonomy shifts of $\lambda_{d-1}$ by the background field $\hat{c}_d$.
In this description, the $(d-1)$-form symmetry is generated by $U(x;k) = \exp (i \, k \,b_0(x))$, which act on the domain walls ${\cal V}(\gamma_{d-1}) = \exp(i \int_{\gamma_{d-1}} \lambda_{d-1} )$.

Following the standard SymTFT-playbook, the $d$-dimensional theory with $(-1)$-form symmetry is obtained with Dirichlet boundary conditions $|\mathfrak{B}^{(\hat{b})} \rangle$ for the $U$ operators, which fixes $b_0 = \hat{b}$ after interval reduction.
Here the fixed background value $\hat{b} \in \frac{2\pi}{N} \mathbb{Z}_N$ plays the role of labeling the $d$-dimensional universe.
The analogous symmetry boundary action,
\begin{align}\label{eq:sym-action_for_universe}
    |\mathfrak{B}^{(\hat{b})} \rangle : \quad S_\text{sym} = -\frac{i N}{2\pi} \int_{\mathfrak{B}_\text{sym}} \hat{b} \wedge c_d = - i \frac{N \hat{b}}{2\pi} \int_{\mathfrak{B}_\text{sym}} c_d\, ,
\end{align}
has the crucial difference to \eqref{eq:sym-action_for_decomp} in that the term analogous to $d\lambda_{d-1}$ is absent, since there are no small gauge transformations for a scalar gauge potential $b_0$.
What will be relevant to us is that in the path-integral for the full $(d+1)$-dimensional system, this boundary action behaves exactly the same as an insertion of the codimension-1 operator
\begin{align}
    V({\mathfrak{B}_\text{sym}}; \tfrac{N\hat{b}}{2\pi}) = \exp\left(i \frac{N\hat{b}}{2\pi} \int_{\mathfrak{B}_\text{sym}} c_d \right) \, .
\end{align}

This allows us to reinterpret the $(-1)$-form symmetry action as insertions of the operator $V(\mathfrak{B}_\text{sym};\frac{N(\hat{b}_2 -\hat{b}_1)}{2\pi})$.
The symmetry transformation which shifts $\hat{b}_1 \rightarrow \hat{b}_2$ is a consequence of the fusion rule of $V$, which in this simple example is just ``adding exponents'',
\begin{align}
    V(\mathfrak{B}_\text{sym}; \tfrac{N \hat{b}_1}{2\pi}) V(\mathfrak{B}_\text{sym};\tfrac{N(\hat{b}_2 -\hat{b}_1)}{2\pi}) = V(\mathfrak{B}_\text{sym};\tfrac{N\hat{b}_2}{2\pi}) \, .
\end{align}
From the $d$-dimensional perspective, this agrees with the intuition that a $(p=-1)$-form symmetry is the insertion of a spacetime-filling topological defect.
While trivially true in this simple example, we shall see later that this is equivalent to stacking the QFT with the worldvolume dynamics of such a defect, which in general can be a fully-fledged TQFT instead of a simple SPT phase.

\subsection{Decomposition and projection operator from the SymTFT}

In this subsection we make contact to the previous literature on decomposition (see, e.g., \cite{Hellerman:2006zs, Sharpe:2014tca, Sharpe:2019ddn, Tanizaki:2019rbk, Komargodski:2020mxz, Cherman:2020cvw, Robbins:2020msp, Nguyen:2021naa, Sharpe:2021srf, Cherman:2021nox, Sharpe:2022ene}), and show specifically how two key features of a theory $\widetilde{\cal T}$ with $(d-1)$-form symmetry --- the decomposition of the partition function into a sum of partition functions, and the existence of topological point operators which act as projectors --- arise naturally from the SymTFT.

The starting point is that the change of boundary conditions in the SymTFT framework ought to be equivalent to the emergence of a quantum-dual $(d-p-2)$-form $\widehat{G}$ symmetry after gauging a $p$-form $G$ symmetry.
Let us demonstrate it explicitly for $p=d-1$ and $G = \hat{G} =  \mathbb{Z}_N$, starting from the boundary conditions \eqref{eq:sym-action_for_decomp}.
For a $(d-1)$-form symmetry, the classification of the possible discrete torsion parameters is simply $H^1(\Z_N,U(1))\cong\Z_N$.
In the Lagrangian description, it is easy to see that, since the gauge field for the $(d-1)$-form symmetry is a top-form $c_d$, the only local term that can be added is proportional to $c_d$.
So we see that, in the case of gauging $(d-1)$-form symmetries, the possible discrete torsion parameters that determine the theory after gauging are in one-to-one correspondence with the possible Dirichlet boundary conditions for the point operators / scalar field in the SymTFT.

First, this gauging procedure can be made explicit in the $d$-dimensional description by starting with \eqref{eq:general_decomposing_lagrangian}, with $\hat{c}_d$ being the background gauge field for the $(d-1)$-form symmetry, and add the discrete torsion term,
\begin{align}\label{eq:gauging_from_lagrangian_ZN}
    S^{(d)} \supset \frac{i N}{2\pi} \int_{{\cal M}_d} b_0 \wedge \left(d \lambda_{d-1} - \hat{c}_d \right) + \frac{i N}{2\pi} \, \hat{b} \int_{{\cal M}_d} \hat{c}_d \, . 
\end{align}
Gauging the $(d-1)$-form symmetry means to promote $\hat{c}_d$ to a dynamical field which we denote by removing the hat.
This leads to the following equations of motion:
\begin{equation}
    \begin{aligned}
        b_0:  && d \lambda_{d-1} =& \, c_d \, , \\
        \lambda_{d-1}: && db_0= & 0 \, ,\quad N \tfrac{b_0}{2\pi} = \, 0 \mod \mathbb{Z} \, , \\
        c_d : && b_0 =& \, \hat{b} \, .
    \end{aligned}
\end{equation}
The last equation precisely determines the universe via the fixed value of $b_0 = \hat{b}$.

In the SymTFT, the two sets of boundary conditions --- Dirichlet for $b_0$ / Neumann for $c_d$ ($|\mathfrak{B}^{(\hat{b})}_\text{sym}\rangle$), and Dirichlet for $c_d$ / Neumann for $b_0$ ($|\mathfrak{B}^{(\hat{c})}_\text{sym}\rangle$) --- are related to each other by Fourier transformation (see, e.g., \cite{Kaidi:2022cpf}),\footnote{We have chosen the normalizations for later convenience; different normalizations correspond to different Euler counterterms for the $d$-dimensional theory which do not affect the results.}
\begin{align}
    | \mathfrak{B}_\text{sym}^{(\hat{c}_d)} \rangle & = \hphantom{\frac{1}{ | H^{0}(\mathfrak{B}_\text{sym}; \mathbb{Z}_N) | }} \sum_{\tfrac{Nb_{0}}{2\pi} \in H^{0}(\mathfrak{B}_\text{sym}; \mathbb{Z}_N) } \exp\left(-\tfrac{i N}{2\pi} \int_{\mathfrak{B}_\text{sym}} b_{0} \wedge \hat{c}_{d} \right) |\mathfrak{B}_\text{sym}^{(b_0)} \rangle \, , \label{eq:fourier_b-to-c} \\
    | \mathfrak{B}_\text{sym}^{(\hat{b}_0)} \rangle & = \frac{1}{ | H^{d}(\mathfrak{B}_\text{sym}; \mathbb{Z}_N) | } \sum_{\tfrac{N c_d}{2\pi} \in H^{d}(\mathfrak{B}_\text{sym}; \mathbb{Z}_N) } \exp\left(\tfrac{i N}{2\pi} \int_{\mathfrak{B}_\text{sym}} \hat{b}_0 \wedge c_{d} \right) |\mathfrak{B}_\text{sym}^{(c_d)} \rangle \, , \label{eq:fourier_c-to-b} 
\end{align}
Given a physical boundary $|\mathfrak{B}_\text{phys}\rangle$ that encodes the (in general non-topological) degrees of freedom of the $d$-dimensional QFT on ${\cal M}_d \cong \mathfrak{B}_\text{phys} \cong \mathfrak{B}_\text{sym}$, this reproduces the gauging-relationship,
\begin{align}
         \widetilde{Z}[\hat{c}_d] & = \langle \mathfrak{B}_\text{phys} \, | \, \mathfrak{B}_\text{sym}^{(\hat{c}_d)} \rangle = \hphantom{\frac{1}{| H^{0}({\cal M}_d; \mathbb{Z}_N)|}} \sum_{\tfrac{Nb_{0}}{2\pi} \in H^{0}} \exp\left(-\tfrac{i N}{2\pi} \int_{{\cal M}_d} b_{0} \wedge \hat{c}_{d} \right) Z[b_0] \, , \label{eq:gauging_b-to-c} \\
         Z[\hat{b}_0] &= \langle \mathfrak{B}_\text{phys} \, | \, \mathfrak{B}_\text{sym}^{(\hat{b}_0)} \rangle = \frac{1}{| H^{d}({\cal M}_d; \mathbb{Z}_N)|} \sum_{\tfrac{Nc_{d}}{2\pi} \in H^{d} } \exp\left(\tfrac{i N}{2\pi} \int_{{\cal M}_d} \hat{b}_{0} \wedge c_{d} \right) \widetilde{Z}[c_d] \, , \label{eq:gauging_c-to-b}
\end{align}
where $\widetilde{Z}[\hat{c}_d]$ is partition function of the global variant $\widetilde{\cal T}$ that has a $(d-1)$-form symmetry coupled to the background field $\hat{c}_d$, and $Z[\hat{b}_0]$ that of the variant ${\cal T}$ with $(-1)$-form symmetry coupled to the background field $\hat{b}_0 \equiv \hat{b}$.
Assuming that the $d$-dimensional spacetime is connected, i.e., $H^0({\cal M}_d;\mathbb{Z}_N) \cong H^d({\cal M}_d; \mathbb{Z}_N) = \mathbb{Z}_N$, these standard SymTFT relations reproduce the salient features of decomposition as follows.

First, the $d$-dimensional theory $\widetilde{\cal T}$ with the $G = \mathbb{Z}_N$ $(d-1)$-form symmetry is known to decompose into $N$ universes ${\cal T}_i$, with a hallmark being that the partition function also decomposes into a sum, $\widetilde{Z} = \sum_i Z_i$ \cite{Hellerman:2006zs, Sharpe:2014tca, Sharpe:2019ddn, Komargodski:2020mxz, Cherman:2020cvw, Robbins:2020msp, Nguyen:2021naa, Sharpe:2021srf, Cherman:2021nox, Sharpe:2022ene} (in these references the background field $\hat{c}_d$ of the $(d-1)$-form symmetry is generically turned off).
This is clearly precisely \eqref{eq:gauging_b-to-c}, where the label $i$ of the universe is identified with different ``configurations'' $b_0 \in \mathbb{Z}_N = H^0 ({\cal M}_d; \mathbb{Z}_N)$ of the $(-1)$-form symmetry's background gauge field, and $Z_i \equiv Z_{b_0} \sim Z[b_0]$ in a trivial background $\hat{c}_d = 0$.
When $\hat{c}_d \neq 0$, the SymTFT conveniently encodes the difference between the universes in terms of the phase factor $\exp( \frac{iN}{2\pi} \int b_0 \, \hat{c}_d) =: \exp( i \, b_0 \, \hat\nu)$, where $\hat\nu = \frac{N}{2\pi} \int_{{\cal M}_d} \hat{c}_d \in \mathbb{Z}_N \cong H^d({\cal M}_d; \mathbb{Z}_N)$ is an equivalent characterization of $\hat{c}_d$.
This factor can be regarded as a discrete torsion for the $(-1)$-form symmetry that distinguishes different universes in the decomposition.

The second equation, \eqref{eq:gauging_c-to-b}, is obviously the same as the gauging \eqref{eq:gauging_from_lagrangian_ZN} which ``undoes'' decomposition and projects onto the universe labeled by $\hat{b}_0$, with the discrete torsion that ``selects'' the universe encoded again in the exponential.
But the SymTFT framework also naturally incorporates the topological point operators $\Pi_{\hat{b}_0}$ which project correlation functions of $\widetilde{\cal T}$ onto their contributions from the universe labeled by $\hat{b}_0$ \cite{Sharpe:2019ddn}.
To see this explicitly, we use the SymTFT's operator/state structure which in this simple case implies
\begin{align}
    U |\mathfrak{B}_\text{sym}^{(c_d)} \rangle \equiv \exp(i b_0) |\mathfrak{B}_\text{sym}^{(\nu)} \rangle = |\mathfrak{B}_\text{sym}^{(\nu+1)} \rangle \, ,
\end{align}
where we have again used $\nu = \frac{N}{2\pi} \int_{\mathfrak{B}_\text{sym}} c_d \in \mathbb{Z}_N$ as an equivalent way of encoding $c_d|_{\mathfrak{B}_\text{sym}}$.
This can be seen, e.g., from \eqref{eq:U_shift_c-periods}, or from the fact that $|\mathfrak{B}_\text{sym}^{(c_d)} \rangle$ form eigenbasis of the SymTFT's Hilbert space for the operator $V$, which in turn is canonically conjugate to $U$.
Then we may rewrite the Fourier transformation \eqref{eq:fourier_c-to-b} as
\begin{align}
\begin{split}
    |\mathfrak{B}_\text{sym}^{(\hat{b}_0)} \rangle & = \frac{1}{|H^d(\mathfrak{B}_\text{sym};\mathbb{Z}_N)|} \sum_{\nu \in \mathbb{Z}_N} \exp( \tfrac{iN}{2\pi} \, \hat{b}_0 \, \nu ) | \mathfrak{B}_\text{sym}^{(\nu)} \rangle = \sum_{\nu \in \mathbb{Z}_N} \frac{\exp( \tfrac{iN}{2\pi} \, \hat{b}_0 \, \nu )}{N} \, U^{-\nu} \, | \mathfrak{B}_\text{sym}^{(c_d \equiv 0)} \rangle \\
    & = \left(\frac{1}{N} \sum_{\nu \in \mathbb{Z}_N} \exp( \tfrac{iN}{2\pi} (\hat{b}_0 - b_0) \nu ) \right) | \mathfrak{B}_\text{sym}^{(c_d \equiv 0)} \rangle =: \Pi_{\hat{b}_0} \, |\mathfrak{B}_\text{sym}^{(c_d \equiv 0)} \rangle \, ,
\end{split}
\end{align}
where
\begin{align}
    \Pi_{\hat{b}_0} := \sum_{\nu=0}^{N-1} \frac{e^{i (\hat{b}_0 - b_0) \nu}}{N} \cong \delta_{b_0, \hat{b}_0}
\end{align}
indeed behaves as a projection operator (i.e., squaring to itself).
When sandwiched with the physical boundary $\langle \mathfrak{B}_\text{phys} |$, this reproduces the partition function of the universes as an expectation value of the projector,
\begin{align}
    Z[\hat{b}_0] = \langle \Pi_{\hat{b}_0} \rangle_{\widetilde{\cal T}_{c_d \equiv 0}} \, ,
\end{align}
in the decomposing theory with \emph{trivial} background for the $(d-1)$-form symmetry. 
Note that by writing $|\mathfrak{B}_\text{sym}^{(c_d \equiv 0)} \rangle = U^{-\hat{\nu}} |\mathfrak{B}_\text{sym}^{(\hat{c}_d)} \rangle$ with $2\pi \hat{\nu} = \int \hat{c}_d$, we can also straightforwardly find the projection operator that projects onto the $\hat{b}_0$-universe from the decomposing theory in any non-trivial background $\hat{c}_d$.

\subsection{\texorpdfstring{$U(1)$ $(-1)$}{U(1) (-1)}-form and  \texorpdfstring{$\mathbb{Z}$ $(d-1)$}{infinite (d-1)}-form symmetries}
\label{ssec:decomp_from_Z}

Using the generalization of the BF-SymTFT to describe $U(1)$ and $\mathbb{Z}$ global symmetries \cite{Antinucci:2024zjp, Brennan:2024fgj, Bonetti:2024cjk, Apruzzi:2024htg}, we can also describe $(-1)$-form symmetries associated with continuous parameter and the corresponding decomposition induced by the dual $(d-1)$-form $\mathbb{Z}$ symmetry.

Following \cite{Antinucci:2024zjp}, we consider the SymTFT action
\begin{align}\label{eq:U(1)/R-SymTFT}
    S = \frac{i}{2\pi} \int_{{\cal W}_{d+1}} d b_0 \wedge c_d \, .
\end{align}
Here $b_0 \simeq b_0 + 2\pi$ is a $U(1)$-valued scalar field with shift symmetry $\delta b_0 = \text{const.}$, and whose periodicity ensures integral periods of its field strength (holonomies), $\frac{1}{2\pi} \oint d b_0 \in \mathbb{Z}$.
Meanwhile, $c_d$ is an $\mathbb{R}$-valued gauge field, which in particular means that it is a globally well-defined differential form with $\oint dc_d = 0$.  The sum over holonomies of $b_0$ in the path integral forces $c_d$ to have integral periods, $\oint c_d\in 2\pi\Z$.
This gives rise to the topological operators
\begin{align}
    \begin{split}
        V(\Gamma_d; \alpha) & = \exp\left( i \alpha \oint_{\Gamma_d} c_d \right) \, , \quad \alpha \in \mathbb{R}/\mathbb{Z} \cong U(1) \, , \\
        U(x ; m) & = \exp(i m b_0(x) ) \, , \quad m \in \mathbb{Z} \, .
    \end{split}
\end{align}
Dragging an operator $U(x;m)$ through $V(\Gamma_d;\alpha)$ results in a braiding phase $\exp(2\pi i \alpha m)$.

Just as the $\mathbb{Z}_N$ case, this SymTFT has two canonical topological boundary conditions on the symmetry boundary $\mathfrak{B}_\text{sym}$ that are analogous.
\begin{enumerate}
    \item By giving Dirichlet b.c.'s to $U$ and Neumann b.c.'s to $V$, we will have a $d$-dimensional theory with $U(1)^{(-1)}$ symmetry after interval reduction.
    This symmetry is generated by insertions of, or stacking the boundary with $V(\mathfrak{B};\alpha)$.
    This shifts the value of $b_0$, which is set to a constant $\hat{b}$, by $2\pi \alpha$.

    The corresponding symmetry boundary action is the straightforward generalization of \eqref{eq:sym-action_for_universe},
    \begin{align}\label{eq:sym-action_for_universe-U(1)}
        |\mathfrak{B}_\text{sym}^{(\hat{b})} \rangle : \quad S_\text{sym} = \frac{-i}{2\pi} \int_{\mathfrak{B}_\text{sym}} \hat{b} \wedge c_d = \frac{-i \, \hat{b}}{2\pi} \int_{\mathfrak{B}_\text{sym}} c_d  \, ,
    \end{align}
    which, when added to \eqref{eq:U(1)/R-SymTFT}, sets $b_0 |_{\mathfrak{B}_\text{sym}} = \hat{b}$.
    Analogous to the finite symmetry case, this boundary condition is equivalent to inserting the operator $V(\mathfrak{B}_\text{sym}; \hat{b})$, and changing the boundary value of $b_0$ --- or, transitioning between universes --- is achieved by further stacking these spacetime-filling operators.

    \item By implementing Dirichlet b.c.~for $c_d$ and Neumann b.c.~for $b_0$, we obtain a $\mathbb{Z}^{(d-1)}$ symmetry with generators $U(x;m)$, which act on $(d-1)$-dimensional operators ${\cal V}(\gamma_{d-1};\alpha)$ which are the $V(\Gamma_d;\alpha)$ which end perpendicularly on $\gamma_{d-1} = \partial \Gamma_d \subset \mathfrak{B}_\text{sym}$.

    Analogously, this corresponds to the SymTFT state $|\mathfrak{B}_\text{sym}^{\hat{c}_d} \rangle$ described by the boundary action
    \begin{align}\label{eq:sym-action_for_decomp-U(1)}
        |\mathfrak{B}_\text{sym}^{(\hat{c}_d)} \rangle : \quad S_\text{sym}[\lambda_{d-1}] = \frac{i}{2\pi} \int_{\mathfrak{B}_\text{sym}} b_0 \wedge (d \lambda_{d-1} - \hat{c}_d) \, ,
    \end{align}
    which together with \eqref{eq:U(1)/R-SymTFT} sets $c_d |_{\mathfrak{B}_\text{sym}} = \hat{c}_d - d\lambda_{d-1}$.
    Here, $\lambda_{d-1}$ is an ordinary differential form which captures the gauge transformations of an $\mathbb{R}$ gauge field, and $\hat{c}_d$ is a fixed flat gauge field, which could also be represented as the flux $\hat{c}_d = d a_{d-1}$ of a $U(1)$ gauge field $a_{d-1}$ \cite{Brennan:2024fgj}.
    This $(d-1)$-form symmetry is the quantum-dual symmetry arising from gauging the $U(1)$ $(-1)$-form symmetry \emph{with discrete topology} \cite{Antinucci:2024zjp}.
\end{enumerate}

Similar to the finite case, the absolute $d$-dimensional theory obtained from the first boundary condition choice (Dirichlet for $U$ / $b_0$) is also the result of gauging the $(d-1)$-form $\mathbb{Z}$ symmetry, or, equivalently, a ``flat'' gauging of the $U(1)$ symmetry associated with $a_{d-1}$ \cite{Brennan:2024fgj}.
Just as in the finite case, the possible discrete torsion parameters for this gauging are $\hat{b} \in H^{1}(\mathbb{Z}, U(1)) \cong U(1)$, which is therefore in one-to-one correspondence with the standard Dirichlet boundary condition choices \eqref{eq:sym-action_for_universe-U(1)} of the SymTFT:
\begin{align}\label{eq:Fourier-trafo-U1/Z-projection}
    |\mathfrak{B}_\text{sym}^{(\hat{b})} \rangle \sim \sum_{c_d \in H^{d} \cong \mathbb{Z}} \exp( \frac{i}{2\pi} \int \hat{b} \, c_d ) |\mathfrak{B}_\text{sym}^{(c_d)} \rangle \, .
\end{align}
In the context of decomposition, this operation has a clear physical interpretation as projecting onto one of the infinite universes associated with the $\mathbb{Z}^{(d-1)}$ symmetry labelled by $\hat{b}$.
The transformation in the other direction is now an integral,
\begin{align}\label{eq:Fourier-trafo-U1/Z-decomposition}
    |\mathfrak{B}_\text{sym}^{(\hat{c}_d)} \rangle \sim \int_{U(1)} d b_0 \exp( \frac{i}{2\pi} \int b_0 \, \hat{c}_d ) |\mathfrak{B}_\text{sym}^{(b_0)} \rangle \, ,
\end{align}
and gives the SymTFT-version of the fact that the theory with $\mathbb{Z}^{(d-1)}$-symmetry decomposes into a continuum of Universes.
We will see momentarily how this interpretation arises in a familiar example.

\section{Instanton restriction in 4d Yang--Mills}\label{sec:instanton_4d}

As a concrete example, let us put decomposition in ``instanton-restricted'' 4d Yang--Mills theory \cite{Tanizaki:2019rbk} associated with $G^{(-1)} = \widehat{G}^{(3)} = \mathbb{Z}_N$ and its generalization to the full $G^{(-1)} = U(1)$, $\widehat{G}^{(3)} = \mathbb{Z}$ into a SymTFT description; similar discussions of the finite case, including stringy realizations, have also appeared in \cite{Yu:2024jtk, Najjar:2024vmm, Najjar:2025htp}.

The standard 4d Yang--Mills Lagrangian,
\begin{align}\label{eq:standard_YM}
    S_\text{YM}[\theta] = \frac{1}{2 g^2} \int_{{\cal M}_4} \text{tr}(F \wedge \star F) + \frac{i \theta}{8\pi^2} \int_{{\cal M}_4} \text{tr}(F \wedge F)
\end{align}
exhibits a $U(1)$ $(-1)$-form symmetry that shifts $\theta \simeq \theta + 2\pi$.
For later convenience, it is useful to recall that the partition function\footnote{We will assume that we have picked the ``electric'' polarization for the center 1-form symmetries, so that the gauge group $G$ is simply-connected, with Wilson-line operators charged under the electic $\mathbb{Z}_N$ 1-form symmetry.} takes the form
\begin{align}\label{eq:standard_YM_partition-function}
    Z^\text{YM}[\theta] = \sum_{\nu\in\mathbb{Z}} e^{-i \theta \, \nu} Z_\nu  \, ,
\end{align}
where $Z_\nu$ is the path integral for the kinetic term $\frac{1}{2g^2} \int_{{\cal M}_4} \text{tr}(F \wedge \star F)$ over field configurations $F$ for which $\tfrac{1}{8\pi^2} \int_{{\cal M}_4} \text{tr}(F^2) = \nu$ is fixed.
Note that this sum is not to be thought of as a consequence of decomposition, as in ordinary Yang--Mills different instanton sectors are connected via non-perturbative transitions, and do not constitute disjoint Universes.

\subsection{Modifying the instanton sum}

To obtain an actual decomposing theory, we gauge (a subgroup of) the $(-1)$-form symmetry.
The subgroup-gauging has appeared previously in \cite{Tanizaki:2019rbk} in disguise, and we shall first review these results and put them into the present context.
In that work, the 4d Yang--Mills Langrangian is coupled to a topological sector,
\begin{align}\label{eq:action_gYM}
\begin{split}
    S_\text{gYM}[\theta;p] & = S_\text{YM}[\theta] + S_\text{top}[p] \, , \\
    \text{with} \qquad S_\text{YM}[\theta] & = \frac{1}{2 g^2} \int_{{\cal M}_4} \text{tr}(F \wedge \star F) + \frac{i \theta}{8\pi^2} \int_{{\cal M}_4} \text{tr}(F \wedge F) \, , \\
    S_\text{top}[p] & = i\int_{{\cal M}_4} b_0 \left( \frac{1}{8\pi^2} \text{tr}(F^2) - \frac{p}{2\pi} d \lambda_3 \right) \, ,
\end{split}
\end{align}
where $b_0$ and $\lambda_3$ are $U(1)$ gauge fields that are summed over in the path-integral.
The authors of \cite{Tanizaki:2019rbk} dubbed this the ``generalized Yang--Mills'' theory.
This theory has the same local dynamics as the Yang--Mills theory, but with a modified instanton sum: the equation of motion for $b_0$, 
\begin{align}\label{eq:coupling_instanton_3-form}
    \frac{1}{8\pi^2} \text{tr}(F^2) = \frac{p}{2\pi} d\lambda_3 \, ,    
\end{align}
restricts the instanton number to be $0 \mod p$, so that the partition function is
\begin{align}
    Z^\text{gYM}[\theta ; p] = \sum_{\nu \in \mathbb{Z}} e^{-i \theta p \nu} Z_{p \nu} \, .
\end{align}

The topological sector $(b_0, \lambda_3)$ is nothing but a $\mathbb{Z}_p$ BF-theory in 4d.
Thus the scalar field with $\frac{p}{2\pi} b_0(x) \equiv 0$ gives $(d-1)=3$-form symmetry with generators $U(x) = \exp(i b_0 (x))$ which induces a decomposition.
In fact, this 3-form symmetry is the quantum symmetry of having gauged a $\mathbb{Z}_p \subset U(1)$ subgroup of the $(-1)$-form symmetry by summing over the on-shell $\mathbb{Z}_p$ field configurations $b_0$.
Indeed, we have argued also from the SymTFT perspective that a gauged $\mathbb{Z}_p$ $(-1)$-form symmetry leads to a $d$-dimensional sector of this form, see \eqref{eq:general_decomposing_lagrangian}.

Notice that we can also turn on a non-trivial background field $\hat{c}_4$ for the 3-form symmetry; in the SymTFT this is associated with the boundary condition $|\mathfrak{B}^{(\hat{c})} \rangle$ with boundary action \eqref{eq:sym-action_for_decomp} which leads to the 4d term (see \eqref{eq:general_decomposing_lagrangian})
\begin{align}
    S_\text{top}[p, \hat{c}_4] = i \int_{{\cal M}_4}  b_0 \left( \frac{1}{8\pi^2} \text{tr}(F^2) - \frac{p}{2\pi} ( d\lambda_3 - \hat{c}_4 ) \right) \, .
\end{align}
Since $\frac{p}{2 \pi} \oint_{{\cal M}_4} \hat{c}_4 \equiv \hat{\nu} \in \mathbb{Z}$ can take any value, the $b_0$ equation of motion leads to $ \int \frac{1}{8\pi^2} \text{tr}(F^2) = \frac{p}{2\pi} \int d\lambda_3 - \frac{p}{2\pi} \int \hat{c}_4 \equiv \hat{\nu} \mod p$.
This means that the partition function is now
\begin{align}
    Z^\text{gYM} [\theta; p, \hat{c}_4] = \sum_{\nu \in \mathbb{Z}} e^{-i\theta (p \nu + \hat{\nu})} \, Z_{p\nu + \hat{\nu}} \, ,
\end{align}
i.e., the instanton sectors are still separated by multiples of $p$, but each has an instanton number with a fixed $\hat{\nu} \mod p$.

While it is clear from the previous section that different Universes in this decomposition will be labelled by different values of the locally constant field $b_0$, the relationship of these with respect to the standard \eqref{eq:standard_YM} and generalized \eqref{eq:action_gYM} Yang--Mills theory may be slightly obscure.
To shed more light on these, we project onto the individual universes, i.e., we gauge the 3-form symmetry (with a discrete torsion parameter $\hat{b} \in \mathbb{Z}_p$).
This means that the background field $c_4$ in the Lagrangian
\begin{align}\label{eq:modified_YM-Lag_with_B4}
    {\cal L} = \frac{1}{2g^2} \text{tr}(F \wedge \star F) + i \left[ \frac{\theta}{8\pi^2} \text{tr}(F^2) + b_0 \left( \frac{1}{8\pi^2} \text{tr}(F^2) - \frac{p}{2\pi} (d\lambda_3 - c_4) \right) - \frac{p}{2\pi} \hat{b} \, c_4 \right] \, ,
\end{align}
is now a dynamical field.
The equations of motion are then
\begin{equation}
    \begin{aligned}
        b_0: && \frac{p}{2\pi} c_4 & = \frac{p}{2\pi} d \lambda_3 - \frac{1}{8\pi^2} \text{tr}(F^2) \, , \\
        \lambda_3: && \frac{p \, b_0}{2\pi} &= 0 \mod \Z \, , \\
        c_4: && b_0 & = \hat{b} \, .
    \end{aligned}
\end{equation}

Firstly, a dynamical $c_4$ acts as a Lagrange multiplier which, as expected, fixes the value of $b_0$.
A second consequence of $c_4$ being dynamical is that the equation of motion of $b_0$,
\begin{align}
    \frac{1}{8\pi^2} \text{tr}(F^2) = \frac{p}{2\pi} (d\lambda_3 - c_4) \, ,
\end{align}
no longer fixes the instanton number to have a fixed modulus mod $p$, since both $\lambda_3$ and $c_4$ are now summed over in the path-integral.
Moreover, the topological sector of \eqref{eq:modified_YM-Lag_with_B4} becomes
\begin{align}
    \begin{split}
        \frac{\theta}{8\pi^2} \text{tr}(F^2) - \frac{p}{2\pi} \hat{b} \, c_4 = \frac{\theta + \hat{b}}{8\pi^2} \text{tr}(F^2)  - \frac{p \, b_0}{2\pi} d\lambda_3 = \frac{\theta + \hat{b}}{8\pi^2} \text{tr}(F^2) + N d \lambda_3 \, , \quad N \in \Z \, .
    \end{split}
\end{align}
Since the second term just integrates to $2\pi i \times $integer which is irrelevant in the path-integral, we see that the gauging of the 3-form symmetry effectively reproduces the standard Yang--Mills Lagrangian, ${\cal L} = \frac{1}{2g^2} \text{tr}(F \wedge \star F) + i \frac{\theta'}{8\pi^2} \text{tr}(F^2)$, with no restriction on the instanton number, but with a shifted $\theta' = \theta + \hat{b}$ angle with $\hat{b} \in \frac{2\pi}{p} \mathbb{Z}$.

In the language of decomposition, this means that the generalized Yang--Mills theory \eqref{eq:action_gYM} with theta angle $\theta$ can be thought of as an ``interference''  \cite{Sharpe:2014tca, Sharpe:2022ene} of disjoint union of $p$ standard Yang--Mills theories at theta angles $\theta$, $\theta + 2\pi/p$, $\theta + 4\pi/p$, ..., $\theta + 2\pi (p-1)/p$.
At the level of the partition function, this decomposition reads
\begin{align}\label{eq:gYM_interference}
\begin{split}
    & \frac{1}{p} \sum_{\ell = 0}^{p-1} Z^\text{YM} \big[ \theta + 2\pi \tfrac{\ell}{p} \big] = \frac{1}{p} \sum_{\ell =0}^{p-1} \sum_{\nu \in \mathbb{Z}} e^{-i (\theta + 2\pi \ell /p) \nu} Z_{\nu} \\
    = \, & \sum_{\nu \in \mathbb{Z}} e^{-i \theta \nu} Z_\nu \, \sum_{\ell = 0}^{p-1} \frac{1}{p} \exp \left(- \tfrac{2\pi i \,\ell}{p} \nu \right) = \sum_{\nu \in \mathbb{Z}} e^{-i \theta \nu} Z_{\nu} \, \delta_{\text{gcd}(\nu,p), p} \\
    = \, & \sum_{\nu \in p \mathbb{Z}} e^{-i \theta \nu} Z_{\nu} = Z^\text{gYM}[\theta;p] \, .
\end{split}
\end{align}

With the insights from Section \ref{ssec:decomp_from_Z}, it is straightforward to further generalize ``generalized Yang--Mills'' by topologically gauging the full $U(1)^{(-1)}$ symmetry of \eqref{eq:standard_YM}.
Specifically, after interval-reducing the SymTFT (which we will present momentarily in full detail) with the boundary condition \eqref{eq:sym-action_for_decomp-U(1)} that gives the $(-1)$-form symmetry gauge field $\chi$ Neumann boundary conditions, we will have an action analogous to \eqref{eq:action_gYM},
\begin{align}\label{eq:action_ggYM}
    S_\text{ggYM}[\hat{c}_4] = \int_{{\cal M}_4} \left( \frac{1}{2g^2} \text{tr}(F \wedge \star F) + \frac{i (\theta + \chi)}{8\pi^2} \text{tr}(F^2) - \frac{i}{2\pi} \chi d \lambda_3 + \frac{i}{2\pi} \chi \hat{c}_4 \right) ,
\end{align}
now with $\chi$ being a $U(1)$-valued scalar field (instead of $b_0$ which is $\mathbb{Z}_p$-valued on-shell), and $\lambda_3$ an ordinary differential form, i.e., $\mathbb{R}$-valued gauge field.
Note that we have included a non-trivial (fixed) background $\hat{c}_4$ for the 3-form symmetry, see \eqref{eq:sym-action_for_decomp-U(1)}.
The equations of motion of the scalar $\chi$,
\begin{align}
    \frac{1}{8\pi^2} \text{tr}(F^2) = \frac{1}{2\pi} (d\lambda_3 - \hat{c}_4) \, ,
\end{align}
while looking identical to \eqref{eq:coupling_instanton_3-form}, has drastically different implications due to the different nature of $\lambda_3$ here:
\begin{align}
    \int_{{\cal M}_4} \frac{1}{8\pi^2} \text{tr}(F^2) = \underbrace{\frac{1}{2\pi} \int_{{\cal M}_4} d \lambda_3}_{=0} -\frac{1}{2\pi} \int_{{\cal M}_4} \hat{c}_4 = -\frac{1}{2\pi} \int_{{\cal M}_4} \hat{c}_4 =: - \hat{\nu} \in \mathbb{Z}
\end{align}
is a fixed integer given by the periods of the flat background $\hat{c}_4$ on any closed ${\cal M}_4$.
So the ``generalized$^2$ Yang--Mills'' theory \eqref{eq:action_ggYM} formally has no instanton sum at all!
Indeed, by integrating out $\chi$ the partition function for the action \eqref{eq:action_ggYM} is simply $Z^\text{ggYM}[\theta] = e^{i \theta \hat{\nu}} \, Z_{\hat{\nu}}$, which is one of the summands in \eqref{eq:standard_YM_partition-function}.
In a sense we have turned this formal sum expression of the standard Yang--Mills partition function into a genuine decomposition, i.e., where individual instanton sectors are disjoint universes, at the expense of introducing the $U(1) / \mathbb{R}$ BF-sector $(\chi, \lambda_3)$.
It may also be naturally viewed as the $p \rightarrow \infty$ limit of the $\mathbb{Z}_p$ decomposition of \cite{Tanizaki:2019rbk} discussed above.

Like in the finite symmetry case, we can uncover the individual universes by gauging, i.e., summing over the background field configurations $\hat{c}_4$.
Including the discrete torsion parameter $\hat{\theta} \in U(1)$, the Lagrangian is
\begin{align}
    {\cal L} =  \frac{1}{2g^2} \text{tr}(F \wedge \star F) + \frac{i \theta + \chi}{8\pi^2} \text{tr}(F^2) - \frac{i}{2\pi} \chi d \lambda_3 + \frac{i}{2\pi} (\chi  - \hat\theta) \hat{c}_4 \, ,
\end{align}
which in complete analogy to \eqref{eq:modified_YM-Lag_with_B4} yields the equations of motion
\begin{equation}
    \begin{aligned}
        \chi: && \frac{1}{2\pi} \hat{c}_4 & = \frac{1}{2\pi} d \lambda_3 - \frac{1}{8\pi^2} \text{tr}(F^2) \, , \\
        \lambda_3: &&  d\chi &= 0 \, , \\
        \hat{c}_4: && \chi & = \hat{\theta} \, ,
    \end{aligned}
\end{equation}
where the first equation now leads leads to a sum over all instanton numbers since $\hat{c}_4$ dynamical.
This reveals the universe to be simply a standard Yang--Mills theory with theta angle $\theta + \hat{\theta}$, as expected.
Thus, we again have an interference effect that reflects the decomposition induced by the $\mathbb{Z}^{(3)}$ symmetry, however with the sum being replaced by an integral:
\begin{align}
\begin{split}
    & \int_{U(1)} \frac{d\hat\theta}{2\pi} \, e^{i \hat\theta \hat{v}} \, Z^{\text{YM}}[\theta + \hat\theta] = \int_{U(1)} \frac{d\hat\theta}{2\pi} \sum_\nu e^{i \hat{\theta} \hat{\nu}}  e^{-i (\theta + \hat\theta) \nu} Z_\nu \\
    = & \sum_\nu \int_{U(1)} \frac{d \hat\theta}{2 \pi} e^{i \hat\theta(\hat\nu - \nu)} e^{-i \theta \nu} Z_\nu  = \sum_\nu \delta(\nu-\hat{\nu}) e^{-i \theta \nu} Z_\nu = e^{-i \theta \hat\nu} Z_{\hat\nu} = Z^\text{ggYM}[\theta] \, ,
\end{split}
\end{align}
where $\hat\nu = \frac{1}{2\pi} \int_{{\cal M}_4} \hat{c}_4$.

\subsection{SymTFT for instanton-restricted Yang--Mills}

The SymTFT description of the instanton restriction described above is straightforward with the tools of the previous section.
In the 5d bulk, we have the $U(1)/\mathbb{R}$-BF theory
\begin{align}\label{eq:sym_bulk_YM}
    S_{\chi} = \frac{i}{2\pi} \int_{{\cal W}_5} d\chi \wedge c_4 \, ,
\end{align}
to which we attach the physical boundary $\mathfrak{B}_\text{phys}$ with localized gapless degrees of freedom given by
\begin{align}
    S_\text{phys} = \frac{1}{2g^2} \int_{\mathfrak{B}_\text{phys}} \text{tr}(F \wedge \star F) + \frac{i\theta}{8\pi^2} \int_{\mathfrak{B}_\text{phys}} \text{tr}(F^2) \, .
\end{align}
The two systems are coupled by the ``interaction'' term
\begin{align}\label{eq:inst-chi-Sym-term}
    S_\text{int} = i \int_{\mathfrak{B}_\text{phys}} \frac{\chi}{8\pi^2} \text{tr}(F^2) \, ,
\end{align}
which imposes $\frac{1}{2\pi} c_4|_{\mathfrak{B}_\text{phys}} = \frac{1}{8\pi^2} \text{tr}(F^2)$ by the equations of motion for $\chi$. 

As explained in \cite{Antinucci:2024zjp}, there is a convenient description of gauging only a $\mathbb{Z}_p$ subgroup of the $U(1)$ $(-1)$-form symmetry on the boundary, by giving Dirichlet boundary conditions for $V(\Gamma_4;\frac{n}{p}) = \exp( i \frac{n}{p} \int_{\Gamma_4} c_4)$, $n \in \mathbb{Z}_p$, as well as $U(x;mp) = \exp(i mp \chi(x))$, $m \in \mathbb{Z}$, on $\mathfrak{B}_\text{sym}$.
The action of the free-floating point operators $U(x;k)$ with $k\neq 0 \mod p$ on the domain walls ${\cal V}(\gamma_3; \frac{n}{p})$ are then identical for $k$ and $k+p$, so that they generate a ${\mathbb{Z}_p}^{(3)}$ symmetry.
Analogously, the free-floating volume operators $V(\Gamma_4;\alpha)$ with $\alpha \neq \frac{n}{p} \mod \mathbb{Z}$ are identified for $\alpha$ and $\alpha+\frac{1}{p}$, and generate a $U(1)/\mathbb{Z}_p \cong U(1)$ $(-1)$-form symmetry which shifts the theta angle of \eqref{eq:action_gYM}.
Equivalently, the action \eqref{eq:sym_bulk_YM} can be rewritten as a coupled theory between an $U(1) - \mathbb{R}$ and a $\mathbb{Z}_p$ SymTFT, where the coupling term reflects a mixed $U(1)^{(-1)} - {\mathbb{Z}_p}^{(3)}$ anomaly \cite{Brennan:2024fgj}.
The above boundary condition would then simply correspond to Dirichlet boundary conditions for the $\mathbb{Z}_p$ scalar field, and leads to the ``generalized Yang--Mills'' theory \eqref{eq:action_gYM} after interval reduction.

The SymTFT further allows for a convenient description of the higher-group symmetry found in \cite{Tanizaki:2019rbk}, and straightforward generalization to the case with $\mathbb{Z}^{(3)}$ symmetry.
To see this, we need to include the 1-form symmetries of the Yang--Mills sector in the SymTFT.
This comes with the standard BF-term (assuming, for simplicity, $Z(G) = \Z_N$ being the center),
\begin{align}
    S_\text{1-form}= \frac{i \, N}{2\pi} \int_{{\cal W}_5} B_2 \wedge dC_2 \, ,
\end{align}
where $B_2$ and $C_2$ are the background fields for the electric and magnetic 1-form symmetries, respectively.
The electric background field further ``interacts'' with the theta angle through the term
\begin{align}\label{eq:genYM-5d_anomaly_term}
    S_\text{anom} = \frac{i \, \alpha}{2\pi} \int_{{\cal W}_5} d \chi \wedge B_2 \wedge B_2 \, ,
\end{align}
where the constant $\alpha$ depends on the non-Abelian gauge algebra.
This is the anomaly ``in the space of couplings'' \cite{Kapustin:2014gua,Gaiotto:2014kfa,Gaiotto:2017yup,Cordova:2019jnf,Cordova:2019uob}, between the large gauge transformations, i.e., $2\pi$-periodicity of $\theta$, and the electric 1-form symmetry.
It can be motivated from the boundary-QFT's perspective by the presence of the coupling \eqref{eq:inst-chi-Sym-term} and the fact that the instanton density fractionalizes in the presence of a background gauge field $B_2$.

For concreteness, let us take $G = SU(N)$, in which case the ``anomaly coefficient'' is $\alpha = \frac{N(N-1)}{2}$.\footnote{We work here in the ``continuous'' description with $U(1)$ gauge fields, whose relation to the discrete background gauge field $\tilde{B}_2 \in H^2({\cal W}_5, \bbZ_{N})$ is roughly $N B_2 = \tilde{B}_2$. In terms of $\tilde{B}_2$, the anomaly coefficient is $\alpha = \frac{N-1}{2N}$, see also \cite{Kapustin:2014gua,Gaiotto:2014kfa,Gaiotto:2017yup,Cordova:2019uob}.}.
Then it is straightforward to verify that the combined bulk action,
\begin{align}
\label{eq:5dfullsymTFT}
    S_\text{Sym} = S_\chi + S_\text{1-form} + S_\text{anom} \, ,
\end{align}
is invariant under the gauge transformations, 
\begin{align}
    \begin{split}
        \delta B_2 & = d \Lambda_1 \, , \quad \delta C_2 = d\tilde\Lambda_1 \, , \\
        \delta c_4 & = d \Lambda_3 - (N-1) \left( N B_2 \wedge d \Lambda_1 + \frac{N}{2} d \Lambda_1 \wedge d\Lambda_1 \right) \, .
    \end{split}
\end{align}
If we give $B_2$ and $c_4$ Dirichlet boundary conditions, i.e., fix them on the boundary to be the background gauge fields of the electric 1-form and the 3-form symmetry, respectively, then the modification of the gauge transformations for $c_4$, which includes the gauge parameter of the $B_2$ transformations, is the hallmark of a higher-group symmetry which mixes the electric $\mathbb{Z}_N$ 1-form and the $\mathbb{Z}$ 3-form symmetry.
This is the natural generalization of the higher-group structure originally found in \cite{Tanizaki:2019rbk} (and later in the top-down approaches \cite{Yu:2024jtk,Najjar:2024vmm,Najjar:2025htp}) for the ${\mathbb{Z}_p}^{(3)}$ symmetry.

As a final comment, this higher-group $\mathbf{G}$ can be equivalently understood as the quantum-dual of gauging the $(-1)$-form symmetry in the presence of the anomaly \eqref{eq:genYM-5d_anomaly_term}.
That is, the 1-form and the 3-form symmetry groups can be considered to form a short exact sequence
\begin{align}
    0 \longrightarrow \mathbb{Z}^{(3)} \longrightarrow \mathbf{G} \longrightarrow {\mathbb{Z}_N}^{(1)} \longrightarrow 0 \, .
\end{align}
A consequence of this is that one can always gauge the 3-form symmetry (thereby ``undoing'' the decomposition) \cite{Sharpe:2019ddn}, but one can generally not gauge the electric 1-form symmetry by itself, i.e., have a decomposing $PSU(N) = SU(N)/\mathbb{Z}_N$ Yang--Mills theory.
This fact is also evident from the SymTFT, where the anomaly \eqref{eq:genYM-5d_anomaly_term}, or, equivalently, the equation of motion for $B_2$,
\begin{align}
    NdC_2 + 2\alpha d\chi \wedge B_2 =0  \, ,
\end{align}
obstructs the simultaneous gauging of the $(-1)$-form and electric 1-form symmetries (i.e., giving $\chi$ and $B_2$ Neumann boundary conditions).

\section{SymTFT for trivially acting \texorpdfstring{$(d-2)$}{(d-2)}-form symmetries}
\label{sec:trivially_acting}

A systematic way to construct a decomposing theory is to start with a theory that has trivially acting $(d-2)$-form symmetries \cite{Pantev:2005rh, Robbins:2020msp, Robbins:2021lry, Robbins:2021ibx, Pantev:2022kpl,Pantev:2022pbf, Robbins:2022wlr, Perez-Lona:2025wup}.
These symmetries correspond to topological line operators, and the fact that they act trivially means that they can end on topological point operators. 
Upon gauging the $(d-2)$-form symmetry, the point operators can become genuine topological point operators, indicating the presence of a $(d-1)$-form symmetry, heralding decomposition.  
In 2d this structure is especially easy to understand, and the examples can be somewhat rich since the trivially acting $0$-form symmetry can sit inside a non-Abelian group in general.
More recently, the notion of trivially acting symmetries has been discussed in the context of the club sandwich construction \cite{Bhardwaj:2023bbf}. 
We will review this construction first and then give an independent construction of trivially acting symmetries within the SymTFT.

\subsection{Trivially acting symmetries: missing charges}

The topological operators that appear in the bulk of $(d+1)$-dimensional SymTFT can capture the generalized charges\cite{Bhardwaj:2023ayw} of a $d$-dimensional $\mathcal{G}$-symmetric QFT. Broadly, there are two types of topological operators in the bulk of SymTFT. 
The first class consists of operators that remain parallel to the symmetry boundary and encode the symmetry generators of the theory. Conversely, the second class consists of operators that extend across both boundaries, generating charged operators (either local or extended) of the theory. 
If for some reason an operator from the latter class does not ``end'' on the physical boundary, the corresponding charged operator will be missing from the absolute QFT. 
In this case the symmetry no longer acts faithfully in the $d$-dimensional theory.

\begin{figure}[ht]
    \centering
    \begin{tikzpicture}[scale=0.65]
    
    \fill[blue!30] (0,0) rectangle (2,3); 
    \fill[green!30] (2,0) rectangle (4,3); 
    
    \draw[thick, black] (2,0) -- (2,3);
    
    \draw[thick] (0,0) -- (0,3);
    \draw[thick] (4,0) -- (4,3);
    
    \node at (-0.5,-0.3) {\footnotesize{$\mathfrak{B}_\text{sym}^\mathcal{G}$}}; 
    \node at (4.5,-0.3) {\footnotesize{$\mathfrak{B}_\text{phys}$}}; 
    \node at (2.3,3.3) {\footnotesize{$\mathcal{I}_\phi$}}; 
    \node at (1.0, 1.65) {\footnotesize{$\mathfrak{Z}(\mathcal{G})$}};
    \node at (3.0, 1.65) {\footnotesize{$\mathfrak{Z}(\mathcal{G}')$}};
    
    \node at (5.5,1.5) {$\rightarrow$}; 

    \begin{scope}[shift={(7,0)}] 
        \fill[green!30] (0,0) rectangle (2,3); 
        
        \draw[thick, black] (2,0) -- (2,3);
        
        \draw[thick] (0,0) -- (0,3);
        \draw[thick] (2,0) -- (2,3); 
        
        \node at (-0.5,-0.3) {}; 
        \node at (2.5,-0.3) {\footnotesize{$\mathfrak{B}_\text{phys}$}}; 
        \node at (0,-0.3) {\footnotesize{$\mathfrak{B}_\text{sym}^{\mathcal{G}'}$}}; 
        \node at (1.1, 1.5) {\footnotesize{$\mathfrak{Z}(\mathcal{G}')$}};
    \end{scope}

    \node at (10.2,1.5) {$\rightarrow$}; 

    \draw[thick, black] (11,0) -- (11,3); 
    \node at (11.5,1.5) {$T$}; 
    
    \end{tikzpicture}
    
    \caption{The club sandwich with the respective SymTFT's of $\mathcal{G}$ and $\mathcal{G}'$. Performing an interval compactification on the blue side of the club sandwich produces a $\mathcal{G}'$-symmetric boundary condition for $T$, 
    $\mathfrak{B}_\text{sym}^{\mathcal{G}'} = \mathfrak{B}_\text{sym}^\mathcal{G} \otimes \mathcal{I}_\phi$. Finally, we obtain the theory $T$ where $\mathcal{G'}$ acts faithfully, by reducing the whole club sandwich.}
    \label{fig:combined_club_sandwich}
\end{figure}
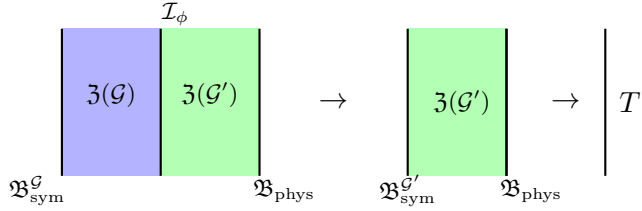

Concretely, we consider a system with a non-faithfully acting symmetry $\mathcal{G}$. 
This is described by a map
\begin{align}
    \Phi: \mathcal{G} \rightarrow \mathcal{G}' \, ,
\end{align}
where the kernel of $\Phi$ is the trivially acting part of $\mathcal{G}$.
Here we should properly view $\mathcal{G}$ and $\mathcal{G}'$ as symmetry categories, and $\Phi$ is a (strong monoidal) functor between them.  In the ordinary group-like case, the restriction of $\Phi$ to objects is simply a group homomorphism, and we will not always be careful in distinguishing between the functor $\Phi$ and this underlying homomorphism.
In a situation like this, we can construct a club sandwich\cite{Bhardwaj:2023bbf, Bhardwaj:2024qrf}, with a topological interface ${\cal I}_\Phi$ between $\mathfrak{Z}(\mathcal{G})$ and $\mathfrak{Z}(\mathcal{G}')$, where $\mathfrak{Z}$ denote the respective SymTFTs, see Fig.~\ref{fig:combined_club_sandwich}.

We would like to review\cite{Bhardwaj:2024qrf} how decomposition manifests in the presence of trivially acting symmetry within the club sandwich picture through a concrete example of 
a QFT $T$ with a $\Z_4$ symmetry, where only a non-anomalous $\Z_{2,e}$ subgroup acts effectively. 
Essentially we have
\begin{equation}
    0 \longrightarrow \Z_{2,t} \longrightarrow \Z_4 \stackrel{\Phi}{\longrightarrow} \Z_{2,e} \longrightarrow 0\, ,
\end{equation}
where $\Z_{2,t}$ is the trivially acting part of $\Z_4$. In this setting, we can construct the club sandwich with the respective SymTFTs\footnote{The interface in the club sandwich is constructed by choosing the condensable algebra $\mathcal{A}_m = 1\oplus m^2$, where we are using the standard nomenclature for the anyons in the Drinfeld center of $\Z_4$, 
\begin{align}\label{5.2}
    \mathcal{Z}(\text{Vec}_{\Z_4}) = \{ e^km^l ~ ~ | ~ ~ k,l \in (0,1,2,3)\} \, . 
\end{align}}. 
The Lagrangian for the 3d SymTFT on the ${\cal G} = \Z_4$ side (cf.~Fig.~\ref{fig:combined_club_sandwich}) is given by, 
\begin{align}
    S^\text{3d}_{\Z_4} = \frac{i}{\pi} \int Ad \hat{A} + Bd \hat{B} -\half Ad \hat{B} \, ,
\end{align}
which is a $\Z_2 \times \Z_2$ Dijkgraaf--Witten theory with a twist. This action is invariant under simple gauge transformations,
\begin{equation}
    \begin{aligned}
        A &\rightarrow A +\delta\lambda \, , \quad B \rightarrow B + \delta \mu \, , \\
        \hat{A} &\rightarrow \hat{A} + \delta \hat{\lambda} \, , \quad \hat{B} \rightarrow \hat{B} +\delta \hat{\mu} \, ,
    \end{aligned}
\end{equation}
and the local equations of motion are simply
\begin{equation}
    \begin{aligned}
    dA =0 \, , \quad d\hat{A} =0 \, , \quad d\hat{B} =0 \, , \quad dB=0 \, .
    \end{aligned}
\end{equation}
In addition to this, the sum over gauge bundle degrees of freedom enforces the following quantization conditions
\begin{equation}
    \begin{aligned}
        2\oint A \in 2\pi \Z \, , \qquad 2\oint (B-\half A) \in 2\pi\Z \, ,\\
        2\oint \hat{B} \in 2\pi \Z \, , \qquad 2\oint (\hat{A} - \half \hat{B}) \in 2\pi\Z \, .
    \end{aligned}
\end{equation}

The most general topological line operators of this theory are given by
\begin{equation}\label{OperatorinZ4}
    \begin{aligned}
        \mathcal{U}_{(n,m,\hat{n},\hat{m})} = \exp [i\oint_{\gamma_1} (nA + mB + \hat{n}\hat{A} + \hat{m}\hat{B})] \, ,
    \end{aligned}
\end{equation}
with the following identifications:
\begin{equation}
    \begin{aligned}
        (n,m,\hat{n},\hat{m}) & \sim (n+2,m,\hat{n},\hat{m}) \, , \\
        (n,m,\hat{n},\hat{m}) & \sim (n,m,\hat{n},\hat{m}+2) \, ,\\
        (n,m,\hat{n},\hat{m}) & \sim (n,m,\hat{n}+2,\hat{m}-1) \, ,\\
        (n,m,\hat{n},\hat{m}) & \sim (n-1,m+2,\hat{n},\hat{m}) \, .
    \end{aligned}
\end{equation}
The fusion rules of the anyons just involve adding the components together and taking the above identifications into account. For example, let us start with $X:=\mathcal{U}_{(0,0,1,0)}= e^{i\oint \hat{A}}$ then, 
\begin{equation}
    \begin{aligned}
    X \otimes X &= \mathcal{U}_{(0,0,2,0)} \\
    & = \mathcal{U}_{(0,0,0,1)} = e^{i\oint \hat{B}} =: U \, ,
    \end{aligned}
\end{equation}
where we have used $(n,m,\hat{n},\hat{m}) \sim (n,m,\hat{n}+2,\hat{m}-1)$. Similarly, we can work out the fusion rule for $V:=\mathcal{U}_{(0,1,0,0)} = e^{i\oint B}$: 
\begin{equation}
    \begin{aligned}
    V \otimes V &= \mathcal{U}_{(0,2,0,0)} \\
                &= \mathcal{U}_{(1,0,0,0)} = e^{i\oint A} =: Y \, ,
    \end{aligned}
\end{equation}
where we have used $(n-1,m+2,\hat{n},\hat{m}) \sim (n,m,\hat{n},\hat{m})$.
We can identify this set $\{X,Y,U,V\}$ of operators with the standard nomenclature of operators in the $\Z_4$-SymTFT \eqref{5.2},
\begin{equation}
    \begin{aligned}
        X = e \, , \quad U=e^2 \, , \quad V=m \, , \quad Y=m^2 \, .
    \end{aligned}
\end{equation}

\begin{figure}[ht]
    \centering
    \begin{tikzpicture}[scale=0.80]
        \begin{scope}
            \fill[blue!30] (0,0,0) -- (2,0,0) -- (2,2,0) -- (0,2,0) -- cycle;
            \fill[blue!15] (0,0,2) -- (0,2,2) -- (2,2,2) -- (2,0,2) -- cycle;
            \fill[blue!25] (0,0,0) -- (0,0,2) -- (2,0,2) -- (2,0,0) -- cycle;
            \fill[blue!25] (0,0,0) -- (0,2,0) -- (0,2,2) -- (0,0,2) -- cycle;
            \fill[blue!30] (0,2,0) -- (2,2,0) -- (2,2,2) -- (0,2,2) -- cycle;
            \fill[blue!30] (2,0,0) -- (2,0,2) -- (2,2,2) -- (2,2,0) -- cycle;

            \fill[green!30] (2,0,0) -- (4,0,0) -- (4,2,0) -- (2,2,0) -- cycle;
            \fill[green!15] (2,0,2) -- (2,2,2) -- (4,2,2) -- (4,0,2) -- cycle;
            \fill[green!25] (2,0,0) -- (2,0,2) -- (4,0,2) -- (4,0,0) -- cycle;
            \fill[green!25] (2,0,0) -- (2,2,0) -- (2,2,2) -- (2,0,2) -- cycle;
            \fill[green!30] (2,2,0) -- (4,2,0) -- (4,2,2) -- (2,2,2) -- cycle;
            \fill[green!30] (4,0,0) -- (4,0,2) -- (4,2,2) -- (4,2,0) -- cycle;

            \draw[thick] (0,0,0) -- (2,0,0) -- (2,2,0) -- (0,2,0) -- cycle;
            \draw[thick] (0,0,0) -- (0,0,2) -- (0,2,2) -- (0,2,0) -- cycle;
            \draw[thick] (0,0,0) -- (2,0,0) -- (2,0,2) -- (0,0,2) -- cycle;
            \draw[thick] (0,2,0) -- (2,2,0) -- (2,2,2) -- (0,2,2) -- cycle;
            \draw[thick] (2,0,0) -- (2,0,2) -- (2,2,2) -- (2,2,0) -- cycle;

            \draw[thick] (2,0,0) -- (4,0,0) -- (4,2,0) -- (2,2,0) -- cycle;
            \draw[thick] (2,0,0) -- (2,0,2) -- (2,2,2) -- (2,2,0) -- cycle;
            \draw[thick] (2,2,0) -- (4,2,0) -- (4,2,2) -- (2,2,2) -- cycle;
            \draw[thick] (4,0,0) -- (4,0,2) -- (4,2,2) -- (4,2,0) -- cycle;
            \draw[thick] (2,0,2) -- (4,0,2) -- (4,2,2) -- (2,2,2) -- cycle;
        
            \node at (1, 0.6, 0.8) {$Y$};
            \node at (-1, 0.9, 0.8) {$\mathcal{L}_\text{sym}^{\Z_4}$};
            \node at (1.5, -0.8, 0.8) {$\mathcal{I}_\phi$};
            \node at (5, 0.9, 0.8) {$\mathfrak{B}_\text{phys}$};

            \draw[thick, blue]  (0,1,1) -- (2,1,1);
            \draw[thick, blue]  (0,1,1) -- (0,0,1);
            \filldraw[blue] (2,1,1) circle (2pt);
            \filldraw[blue] (0,1,1) circle (2pt);
            \end{scope}

            \node at (6,1) {\Huge $\rightarrow$};

            \begin{scope}[shift={(8,-0.8)}]
            \fill[green!20] (0,0) -- (2,0) -- (2,2) -- (0,2) -- cycle;
            \fill[green!15] (0,0) -- (0,2) -- (2,2) -- (2,0) -- cycle;
            \fill[green!25] (0,0) -- (0,0,-2) -- (2,0,-2) -- (2,0) -- cycle;
            \fill[green!25] (0,0) -- (0,2,0) -- (0,2,-2) -- (0,0,-2) -- cycle;
            \fill[green!20] (0,2) -- (2,2) -- (2,2,-2) -- (0,2,-2) -- cycle;
            \fill[green!20] (2,0) -- (2,0,-2) -- (2,2,-2) -- (2,2) -- cycle;

            \draw[thick] (0,0) -- (2,0) -- (2,2) -- (0,2) -- cycle;
            \draw[thick] (0,0) -- (0,0,-2) -- (0,2,-2) -- (0,2) -- cycle;
            \draw[thick] (0,0) -- (2,0) -- (2,0,-2) -- (0,0,-2) -- cycle;
            \draw[thick] (0,2) -- (2,2) -- (2,2,-2) -- (0,2,-2) -- cycle;
            \draw[thick] (2,0) -- (2,0,-2) -- (2,2,-2) -- (2,2) -- cycle;

            \draw[thick, blue] (0.5,1.3,0.3) -- (0,0,-1);
            \filldraw[blue] (0.5,1.3,0.3) circle (2pt);

            \node at (0, 0.9, 0.8) {$\mathcal{L}_\text{sym}^{\Z_2}$};
            \node at (0, 0.9, -1.5) {$\mathcal{Q}_m$};
            \node at (3, 1.9, 1) {$\mathfrak{B}_\text{phys}$};
        \end{scope}

    \end{tikzpicture}
    \caption{We have the club sandwich with the $\Z_4$ and $|Z_2$ SymTFT's. The interface corresponds to a choice of condensable algebra $ \mathcal{A} = 1 \oplus m^2 = 1 \oplus Y$ .The operator $Y$ remains topological on the symmetry boundary. Upon performing an interval compactification on the blue side,  which results in a  $\Z_2$ symmetric boundary condition for $T$, we obtain the lines that terminates on topological point operator, in addition to the lines that generate the $\Z_2$ symmetry.}
    \label{fig:combined_diagram}
\end{figure}
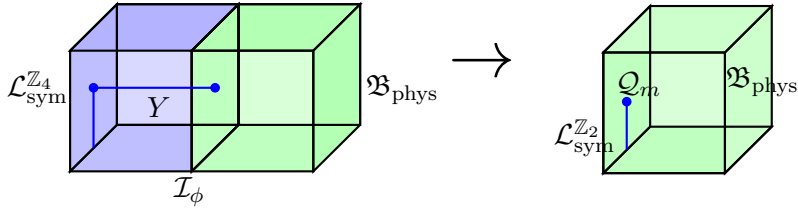

In order to realize the $\Z_4$ symmetry, we must impose Dirichlet boundary condition for $\hat{A}$ and $\hat{B}$ on the symmetry boundary\footnote{This is equivalent to choosing the Lagrangian algebra on the symmetry boundary as $\mathcal{L}^{\Z_4}_\text{sym} = 1 \oplus e^1 \oplus e^2 \oplus e^3$.}. Our primary interest is in the operator $Y$, which remains topological on the symmetry boundary (with the chosen boundary condition), but terminates on the interface. On the other hand, the operators $X$ and $XU$ terminate on the symmetry boundary but cannot reach the physical boundary because they cannot penetrate the interface and ``enter'' the $\Z_2$ side of the club sandwich\footnote{We can do a systematic computation of reduced topological in order to see this, for further details see Appendix B of \cite{Bhardwaj:2023bbf}.}. In a theory with faithfully acting $\Z_4$ symmetry, both $X$ and $XU$ would have terminated on the physical boundary, producing the local operators charged under the $\Z_4$ symmetry. Their inability to reach the physical boundary in the present case indicates that only a $\Z_2$ subgroup of $\Z_4$ acts faithfully, as the charged operators associated with them will be absent from the $\Z_4$ theory. In addition to the missing charges, we also note that when we collapse the blue side of the club sandwich, we observe the presence of topological lines that terminate on topological point operators localized on the $\Z_2$ symmetry boundary, as in Fig.~\eqref{fig:combined_diagram}.  

\subsubsection*{Gauging the trivially acting symmetry}
\begin{figure}[ht]
    \centering
    \begin{tikzpicture}[scale=0.85]
        \fill[blue!25] (0,0,0) -- (2,0,0) -- (2,2,0) -- (0,2,0) -- cycle;
        \fill[blue!15] (0,0,2) -- (0,2,2) -- (2,2,2) -- (2,0,2) -- cycle;
        \fill[blue!20] (0,0,0) -- (0,0,2) -- (2,0,2) -- (2,0,0) -- cycle;
        \fill[blue!25] (0,0,0) -- (0,2,0) -- (0,2,2) -- (0,0,2) -- cycle;
        \fill[blue!30] (0,2,0) -- (2,2,0) -- (2,2,2) -- (0,2,2) -- cycle;
        \fill[blue!30] (2,0,0) -- (2,0,2) -- (2,2,2) -- (2,2,0) -- cycle;
        
        \fill[green!30] (2,0,0) -- (4,0,0) -- (4,2,0) -- (2,2,0) -- cycle;
        \fill[green!15] (2,0,2) -- (2,2,2) -- (4,2,2) -- (4,0,2) -- cycle;
        \fill[green!25] (2,0,0) -- (2,0,2) -- (4,0,2) -- (4,0,0) -- cycle;
        \fill[green!25] (2,0,0) -- (2,2,0) -- (2,2,2) -- (2,0,2) -- cycle;
        \fill[green!30] (2,2,0) -- (4,2,0) -- (4,2,2) -- (2,2,2) -- cycle;
        \fill[green!30] (4,0,0) -- (4,0,2) -- (4,2,2) -- (4,2,0) -- cycle;
        
        \draw[thick] (0,0,0) -- (2,0,0) -- (2,2,0) -- (0,2,0) -- cycle;
        \draw[thick] (0,0,0) -- (0,0,2) -- (0,2,2) -- (0,2,0) -- cycle;
        \draw[thick] (0,0,0) -- (2,0,0) -- (2,0,2) -- (0,0,2) -- cycle;
        \draw[thick] (0,2,0) -- (2,2,0) -- (2,2,2) -- (0,2,2) -- cycle;
        \draw[thick] (2,0,0) -- (2,0,2) -- (2,2,2) -- (2,2,0) -- cycle;
        
        \draw[thick] (2,0,0) -- (4,0,0) -- (4,2,0) -- (2,2,0) -- cycle;
        \draw[thick] (2,0,0) -- (2,0,2) -- (2,2,2) -- (2,2,0) -- cycle;
        \draw[thick] (2,2,0) -- (4,2,0) -- (4,2,2) -- (2,2,2) -- cycle;
        \draw[thick] (4,0,0) -- (4,0,2) -- (4,2,2) -- (4,2,0) -- cycle;
        \draw[thick] (2,0,2) -- (4,0,2) -- (4,2,2) -- (2,2,2) -- cycle;
        
        \draw[thick, black] (2,0,0) -- (2,0,2);
        \draw[thick, black] (2,2,0) -- (2,2,2);
        \draw[thick, black] (4,0,0) -- (4,0,2);
        \draw[thick, black] (4,2,0) -- (4,2,2);
        
        \draw[thick, black] (-0.2,0.8,0) -- (1.6,0.8,0);
        \filldraw[blue] (-0.2,0.8,0) circle (1.8pt);
        \filldraw[blue] (1.6,0.8,0) circle (1.8pt);
        
        
        \node at (-1, 0.7, 0.8) {$\mathcal{L}_\text{sym}^{\hat{\Z_4}}$};
        \node at (2, 0.7, 0.8) {$\mathcal{I}_\phi$};
        \node at (5, 0.7, 0.8) {$\mathfrak{B}_\text{phys}$};
        \node at (1, 0.9, 0.8) {$Y$};

        \node at (6,1) {\Huge $\rightarrow$};

        \begin{scope}[shift={(8,-0.8)}]
            \fill[green!20] (0,0) -- (2,0) -- (2,2) -- (0,2) -- cycle;
            \fill[green!15] (0,0) -- (0,2) -- (2,2) -- (2,0) -- cycle;
            \fill[green!25] (0,0) -- (0,0,-2) -- (2,0,-2) -- (2,0) -- cycle;
            \fill[green!25] (0,0) -- (0,2,0) -- (0,2,-2) -- (0,0,-2) -- cycle;
            \fill[green!20] (0,2) -- (2,2) -- (2,2,-2) -- (0,2,-2) -- cycle;
            \fill[green!20] (2,0) -- (2,0,-2) -- (2,2,-2) -- (2,2) -- cycle;

            \draw[thick] (0,0) -- (2,0) -- (2,2) -- (0,2) -- cycle;
            \draw[thick] (0,0) -- (0,0,-2) -- (0,2,-2) -- (0,2) -- cycle;
            \draw[thick] (0,0) -- (2,0) -- (2,0,-2) -- (0,0,-2) -- cycle;
            \draw[thick] (0,2) -- (2,2) -- (2,2,-2) -- (0,2,-2) -- cycle;
            \draw[thick] (2,0) -- (2,0,-2) -- (2,2,-2) -- (2,2) -- cycle;

            \filldraw[blue] (0.5,1.3,0.3) circle (2pt);

            \node at (0, 0.9, 0.8) {$\mathcal{L}_\text{sym}^{\Z_2}$};
            \node at (0, 0.9, -1.5) {$\mathcal{Q}_m$};
            \node at (3, 1.9, 1) {$\mathfrak{B}_\text{phys}$};
        \end{scope}
    \end{tikzpicture}
    \caption{We have changed the symmetry boundary condition, as a result the operator $Y$, now ``ends'' on both the symmetry boundary as well as on the interface.}
    \label{fig:gauge_trivial_sym}
\end{figure}
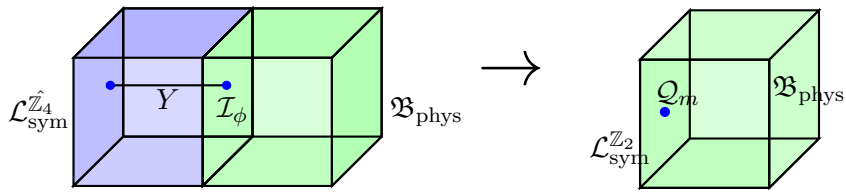

We can now talk about gauging the $\Z_4$ symmetry. In order to do that, we impose Dirichlet boundary condition on $A$ and $B$ as our symmetry boundary condition. This is equivalent to saying that the new symmetry boundary is given by the Lagrangian algebra $\mathcal{L}_\text{sym}^{\hat{\Z_4}}= 1\oplus m \oplus m^2 \oplus m^3$ (alternatively we could gauge just the trivially acting $\Z_2$ subgroup using the Lagrangian algebra $1\oplus m^2\oplus e^2\oplus e^2m^2$; the qualitative discussion will not change much). This will lead us to decomposition. As an immediate consequence, we observe that the $Y$ line now terminates on the symmetry boundary as well as on the interface. If we reduce the blue side of the club sandwich, we obtain the situtation depicted in Fig.~\ref{fig:gauge_trivial_sym}.

We realize the existence of non-trivial topological point operators living on the $\Z_2$ symmetry boundary.
Emergence of these point operators on the $\Z_2$ symmetric boundary condition for our theory $T$ implies that our theory decomposes into universes, as a consequence of gauging the trivially acting 0-form symmetry. 

\begin{figure}[ht]
\centering
\begin{tikzpicture}[scale=0.65]

    \draw[thick] (-2,0) -- (-2,3.3) node[below] at (-2, -0.3) {$\mathfrak{B}^{\Z_4}_\text{sym}$}; 
    \draw[thick] (1,0) -- (1,3.3) node[below] at (1,-0.3) {$\mathcal{I}_\phi$}; 
    \draw[thick] (4,0) -- (4,3.3) node[below] at (4,-0.3) {$\mathfrak{B}_\text{phys}$}; 

    \fill[blue!20] (-2,0) rectangle (1,3.3);

    \draw[thick] (-2.0,2.5) -- (1,2.5) node[midway, above] {$Y$}; 
    \draw[thick] (-2.0,1.0) -- (1,1.0) node[midway, above] {$Y$};

    \filldraw[red] (1,1.0) circle (0.15cm) node[above] at (1.3,1.0) {$\mathcal{O}$};
    \filldraw[red] (1,2.5) circle (0.15cm) node[above] at (1.3,2.5) {$\mathcal{O}$};


    \draw[thick] (13,0) -- (13,3.3) node[below] at (13,-0.3) {$\mathfrak{B}^{\Z_2}_\text{sym}$}; 
    \draw[thick] (16,0) -- (16,3.3) node[below] at (16,-0.3) {$\mathfrak{B}_\text{phys}$}; 

    \filldraw[red] (13,1.0) circle (0.15cm) node[above] at (13.3,1.0) {$\mathcal{O}$};
    \filldraw[red] (13,2.5) circle (0.15cm) node[above] at (13.3,2.5) {$\mathcal{O}$};

\end{tikzpicture}
\caption{Left: The operator $Y$ terminates on the interface. We take two of them and fuse them before they terminate on the interface, producing an identity line that terminates on an identity operator.  
Right: Closing the $\mathbb{Z}_4$ part first produces topological point operators on the $\mathbb{Z}_2$ symmetric boundary. These operators can now be fused. The final result from this process must match that of the left diagram.}
\label{fig:p_op_prod}
\end{figure}
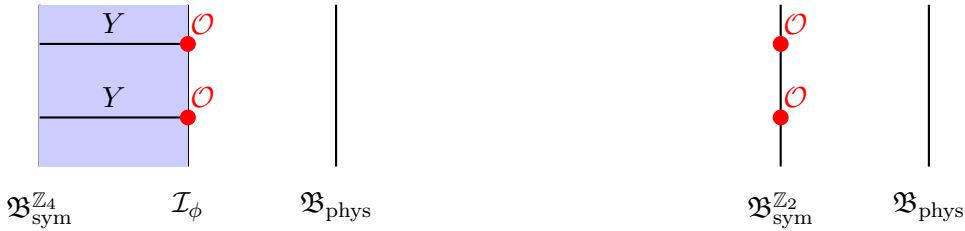

We can identify the product structure of the emerging point operators. The operator $Y$ along with the identity line operator can extend from the interface to the symmetry boundary. The end point of the identity line gives the identity point operator whereas $Y$ terminates on a non-identity topological point operator $\mathcal{O}$. To analyze the fusion, we take two such $Y$ lines before collapsing the trivially acting part of the club sandwich. We then compare it with our result, where we first collapse the club sandwich and then fuse the topological point operators living on the $\Z_2$ symmetric boundary condition (of $T$). 
Using the fact that the $Y$ line squares to identity, $Y \otimes Y = m^2 \otimes m^2 = 1$ and the arguments given in Fig.~\eqref{fig:p_op_prod} allows us to identify the product structure of $\mathcal{O}$, 
\begin{equation}\label{TPOfusion}
    \mathcal{O} \otimes \mathcal{O} = 1
\end{equation}
This allows us to associate the set of point operators $\{ id, \mathcal{O}\}$ with a $\Z_2^{(1)}$ symmetry in the absolute 2d theory, which leads us to decomposition.

\paragraph{An Aside:}
Contrary to popular belief, gauging trivially acting $(d-2)$-form symmetry \textbf{does not always lead to decomposition}. As an example, we can talk about a $\Z_4$ theory with faithfully acting anomalous $\Z^{\omega}_2$ subgroup. 
This is equivalent to saying that, instead of working with the condensable algebra $\mathcal{A}_m$, we work with $\mathcal{A}_{e^2m^2} = 1 \oplus e^2m^2$. 
We will obtain a club sandwich with $\Z_4$ and $\Z_2^\omega$ SymTFTs. 

\begin{figure}[ht]
    \centering
    \begin{tikzpicture}[scale=0.75]
        \fill[blue!30] (0,0,0) -- (2,0,0) -- (2,2,0) -- (0,2,0) -- cycle;
        \fill[blue!15] (0,0,2) -- (0,2,2) -- (2,2,2) -- (2,0,2) -- cycle;
        \fill[blue!25] (0,0,0) -- (0,0,2) -- (2,0,2) -- (2,0,0) -- cycle;
        \fill[blue!25] (0,0,0) -- (0,2,0) -- (0,2,2) -- (0,0,2) -- cycle;
        \fill[blue!30] (0,2,0) -- (2,2,0) -- (2,2,2) -- (0,2,2) -- cycle;
        \fill[blue!30] (2,0,0) -- (2,0,2) -- (2,2,2) -- (2,2,0) -- cycle;
        
        \fill[green!30] (2,0,0) -- (4,0,0) -- (4,2,0) -- (2,2,0) -- cycle;
        \fill[green!15] (2,0,2) -- (2,2,2) -- (4,2,2) -- (4,0,2) -- cycle;
        \fill[green!25] (2,0,0) -- (2,0,2) -- (4,0,2) -- (4,0,0) -- cycle;
        \fill[green!25] (2,0,0) -- (2,2,0) -- (2,2,2) -- (2,0,2) -- cycle;
        \fill[green!30] (2,2,0) -- (4,2,0) -- (4,2,2) -- (2,2,2) -- cycle;
        \fill[green!30] (4,0,0) -- (4,0,2) -- (4,2,2) -- (4,2,0) -- cycle;
        
        \draw[thick] (0,0,0) -- (2,0,0) -- (2,2,0) -- (0,2,0) -- cycle;
        \draw[thick] (0,0,0) -- (0,0,2) -- (0,2,2) -- (0,2,0) -- cycle;
        \draw[thick] (0,0,0) -- (2,0,0) -- (2,0,2) -- (0,0,2) -- cycle;
        \draw[thick] (0,2,0) -- (2,2,0) -- (2,2,2) -- (0,2,2) -- cycle;
        \draw[thick] (2,0,0) -- (2,0,2) -- (2,2,2) -- (2,2,0) -- cycle;
        
        \draw[thick] (2,0,0) -- (4,0,0) -- (4,2,0) -- (2,2,0) -- cycle;
        \draw[thick] (2,0,0) -- (2,0,2) -- (2,2,2) -- (2,2,0) -- cycle;
        \draw[thick] (2,2,0) -- (4,2,0) -- (4,2,2) -- (2,2,2) -- cycle;
        \draw[thick] (4,0,0) -- (4,0,2) -- (4,2,2) -- (4,2,0) -- cycle;
        \draw[thick] (2,0,2) -- (4,0,2) -- (4,2,2) -- (2,2,2) -- cycle;
        
        \draw[thick, black] (2,0,0) -- (2,0,2);
        \draw[thick, black] (2,2,0) -- (2,2,2);
        \draw[thick, black] (4,0,0) -- (4,0,2);
        \draw[thick, black] (4,2,0) -- (4,2,2);
        
        \node at (1, 0.6, 0.8) {$UY$};
        \node at (-1, 0.9, 0.8) {$\mathcal{L}_\text{sym}^{\hat{\Z}_4}$};
        \node at (1.5, -0.8, 0.8) {$\mathcal{I}_\phi$};
        \node at (5, 0.9, 0.8) {$\mathfrak{B}_\text{phys}$};

        \draw[thick, blue]  (0,1,1) -- (2,1,1);
        \draw[thick, blue]  (0,1,1) -- (0,0,1);
        \filldraw[blue] (2,1,1) circle (2pt);
        \filldraw[blue] (0,1,1) circle (2pt);

        \node at (6,1) {\Huge $\rightarrow$};

        \begin{scope}[shift={(7.2,-0.9)}]
            \fill[green!20] (0,0) -- (2,0) -- (2,2) -- (0,2) -- cycle;
            \fill[green!15] (0,0) -- (0,2) -- (2,2) -- (2,0) -- cycle;
            \fill[green!25] (0,0) -- (0,0,-2) -- (2,0,-2) -- (2,0) -- cycle;
            \fill[green!25] (0,0) -- (0,2,0) -- (0,2,-2) -- (0,0,-2) -- cycle;
            \fill[green!20] (0,2) -- (2,2) -- (2,2,-2) -- (0,2,-2) -- cycle;
            \fill[green!20] (2,0) -- (2,0,-2) -- (2,2,-2) -- (2,2) -- cycle;

            \draw[thick] (0,0) -- (2,0) -- (2,2) -- (0,2) -- cycle;
            \draw[thick] (0,0) -- (0,0,-2) -- (0,2,-2) -- (0,2) -- cycle;
            \draw[thick] (0,0) -- (2,0) -- (2,0,-2) -- (0,0,-2) -- cycle;
            \draw[thick] (0,2) -- (2,2) -- (2,2,-2) -- (0,2,-2) -- cycle;
            \draw[thick] (2,0) -- (2,0,-2) -- (2,2,-2) -- (2,2) -- cycle;

            \draw[thick, blue] (0.5,1.3,0.3) -- (0,0,-1);
            \filldraw[blue] (0.5,1.3,0.3) circle (2pt);

            \node at (0, 0.9, 0.8) {$\mathcal{L}_\text{sym}^{\hat{\Z}_2}$};
            \node at (0, 0.9, -1.5) {$\mathcal{Q}_m$};
            \node at (3, 1.9, 1) {$\mathfrak{B}_\text{phys}$};
        \end{scope}
        
    \end{tikzpicture}
    \caption{Even after gauging, we can see that the operator $UY$ does not terminate on the symmetry boundary. When we close the sandwich, we do not get any free point operators on the symmetry boundary.} 
    \label{fig:gauging_trivial_sym_em2}
\end{figure}
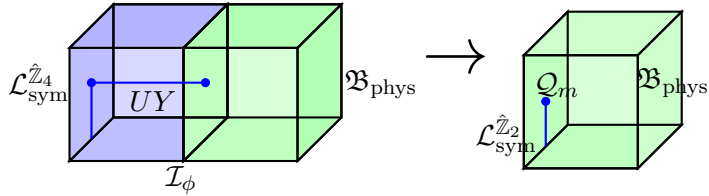

In this set up, the $e^2m^2$ anyon (which is equivalent to $U \otimes Y = e^{i\int (A + \hat{B})}$) terminates on the interface. Our symmetry boundary condition remains unchanged, that is, we impose Dirichlet b.c.~for $\hat{A}$ and $\hat{B}$ and Neumann b.c.~for $A,B$. Therefore, the operator $U$ terminates on the symmetry boundary while $Y$ does not. Like before, we gauge the symmetry by changing the boundary condition. In doing so, we expect to see point operators once we collapse the blue side of the club sandwich. Surprisingly, this does not happen in this case because the operator $U\otimes Y$ cannot terminate on the symmetry boundary. Although the prescribed boundary condition (Dirichlet b.c.~for $A$ and $B$) ensures that $Y$ terminates on the symmetry boundary, $U$ must continue on parallel to the boundary. That is why after collapsing the $\Z_4$ part of the club sandwich, we do not realize \textit{free} topological point operators on the $\Z_2$ symmetric boundary condition. Instead, we end up with a situation similar to Fig.~\ref{fig:gauging_trivial_sym_em2}.

\subsection{Trivially acting symmetries: trivial symmetry generators}

In the previous subsection, we discussed the notion of trivially acting symmetries from the club sandwich. In this section, we would like to give an independent construction of these non-faithfully acting symmetries without evoking the club sandwich picture.
The club sandwich picture relied on the concept of missing charges to describe the non-faithfully acting symmetries. There is another way in which we can approach the trivially acting symmetries, based on trivially acting symmetry generators. In 2d these correspond to topological line operators that end on topological point operators.  The fact that these lines can end on points means that we can always ``unhook'' the line from any linking, so in particular nothing can be charged under the corresponding symmetry\footnote{For an illustration of this, see \cite[Fig.~6]{Robbins:2022wlr}.}.

Let us try to construct a SymTFT that can describes this sort of situation. We begin with the case where the symmetries involved are all $\Z_2$.  We need a dual pair of $U(1)$ background fields, $A_1$ and $\hat{A}_1$.  They each have gauge transformations, $A_1\rr A_1+d\lambda_0$ and $\hat{A}_1\rr\hat{A}_1+d\hat{\lambda}_0$, and an action in the 3D SymTFT,
\begin{equation}
    S\supset\frac{i}{\pi}\int\hat{A}_1dA_1 \, .
\end{equation}
The electric $\Z_2$ $0$-form symmetry is generated by lines $\hat{W}(M_1)=\exp(i\oint_{M_1}\hat{A}_1)$, while the magnetic $\Z_2$ $0$-form symmetry is generated by the lines $W(M_1)=\exp(i\oint_{M_1}A_1)$.

If we claim that the electric symmetry acts trivially, then it should be that the $\hat{W}$ lines can end.  To accomplish this, let us also introduce a background field $B_0$ and its magnetic dual $C_2$, appearing in the action as a term $\frac{i}{\pi}\int C_2d B_0$.  We give $C_2$ a standard $1$-form symmetry gauge transformation, $C_2\rr C_2+d\mu_1$, and in order to make the $\hat{W}$ lines endable, we propose that $B_0$ should transform as $B_0\rr B_0-\hat{\lambda}_0$.  In that case, we can write down gauge invariant topological operators associated to a curve $C_1$ that goes from a point $x$ to a point $y$,
\begin{equation}
    \hat{W}(C_1)=\exp\left(i\int_x^y\hat{A}_1\right)\exp(iB_0(y))\exp(-iB_0(x)) \, .
\end{equation}

This gives us the trivially acting symmetry that we wanted, but now our action is no longer gauge invariant under the $\hat{\lambda}_0$ transformation,
\begin{equation}
S\rightarrow S-\frac{i}{\pi}\int C_2d\hat{\lambda}_0 \, .
\end{equation}
To fix this, we can add another term to the action, giving us
\begin{equation}
\label{eq:DecompToyS}
    S=\frac{i}{\pi}\int\left(\hat{A}_1d A_1+C_2d B_0+C_2\hat{A}_1\right) \, .
\end{equation}
The third term can perhaps be interpreted as a mixed anomaly between the magnetic $0$-form symmetry and the $1$-form symmetry whose background field is $C_2$.  This action is now invariant under the $\hat{\lambda}_0$ transformations, but we now have a problem with the $\mu_1$ transformation.  But this can also be fixed if we propose that $A_1$ shifts under such a gauge transformation.  That is, in summary, we postulate the following gauge transformations to go with the action \eqref{eq:DecompToyS},
\begin{equation}
\begin{aligned}
    A_1\rr\ & A_1+d\lambda_0-\mu_1 \,,\\
    \hat{A}_1\rr\ & \hat{A}_1+d\hat{\lambda}_0 \,,\\
    B_0\rr\ & B_0-\hat{\lambda}_0 \,,\\
    C_2\rr\ & C_2+d\mu_1 \,.
\end{aligned}
\end{equation}
The equations of motion are, 
\begin{equation}\label{EQMTrivActSym}
\begin{aligned}
    d\hat{A}_1=\ & 0 \, , \qquad d C_2=\  0 \, ,\\
    d A_1+C_2=\ & 0 \, , \qquad d B_0+\hat{A}_1=\  0 \, ,
\end{aligned}
\end{equation}
but note that the first two equations actually follow from the last two, so only these last two are independent.

What about the gauge-invariant operators in the SymTFT bulk?  The only ones we have are the previously constructed line operators,
\begin{equation}\label{triviallines}
    \hat{W}(C_1)=\exp\left(i\int_x^y\hat{A}_1\right)\exp(iB_0(y))\exp(-iB_0(x)) \, ,
\end{equation}
with the loop operators $\hat{W}(M_1)=\exp(i\oint_{M_1}\hat{A}_1)$ as a special case. These are topological lines that terminate on topological point operators. We obtained these non-genuine point operators (i.e., the point operators can only exist when attached to lines) directly from the SymTFT action. The information about these operators is essentially captured by the extra couplings in the SymTFT action.  We also have surface operators which can terminate on lines,
\begin{equation}
    V(\Sigma_2)=\exp\left(i\int_{\Sigma_2}C_2\right)\exp\left(i\oint_{\partial\Sigma_2}A_1\right),
\end{equation}
with closed surface operators $V(M_2)=\exp(i\oint_{M_2}C_2)$ as a special case.

Now we want to think about possible topological boundary conditions.  Because of the two $BF$-type terms in the action, we cannot have Neumann conditions for both $A_1$ and $\hat{A}_1$, nor for both $B_0$ and $C_2$, and because of the mixed anomaly term, we cannot have Neumann simultaneously for both $C_2$ and $\hat{A}_1$.  This leaves three interesting cases (see Table \ref{tab:summaryoftas} for a brief summary).

\begin{itemize}
    \item[1)] We take Dirichlet for $A_1$ and $C_2$, and take Neumann for $\hat{A}_1$ and $B_0$.  Then we can put our trivially-acting line operators $\hat{W}(C_1)$ parallel to the symmetry boundary, and so our theory will have these trivial symmetries. Since these lines terminate on points, they usually do not link with any other defect by construction. This manifests its trivially acting nature. The chosen boundary condition also rules out free topological point operators, eliminating the possibility of having a 1-form symmetry and decomposition.  
    
    \item[2)] We take Dirichlet for $\hat{A}_1$ and $C_2$, Neumann for $A_1$ and $B_0$.  This should correspond to gauging the trivially-acting electric $\Z_2$ symmetry. Now, the bulk operators $V(\Sigma_2)$ can end on a line operator $\exp(i\oint_{M_1}A_1)$ on the boundary, and the bulk $\hat{W}(C_1)$ operators can end on a point operator $e^{iB_0(x)}$ on the boundary.  Although both of these operators must attach to operators extended into the bulk, they can remain topological because the bulk operators do not have to extend all the way to the dynamical boundary; the surface $\Sigma_2$ can simply cap off (if $M_1$ is contractible in the bulk), or the curve $C_1$ can end at a point in the bulk. We can also have $V(\Sigma_2)$ stretching all the way to the physical boundary.  We interpret this case as exhibiting decomposition with topological point operators generating a $1$-form symmetry, and with the boundary line operators acting as domain walls between universes. This story is also consistent with our earlier discussion of trivially acting symmetries using the club sandwich, where we saw that gauging the trivially acting symmetry produced topological point operators on the symmetry boundary. Here we observed in a different way how gauging the trivially acting symmetry frees up the topological point operators. We end up with a 1-form symmetry and a quantum-dual 0-form symmetry. 
    
    \item[3)] Finally, we could take Dirichlet for $\hat{A}_1$ and $B_0$, and Neumann for $A_1$ and $C_2$.  This should correspond to starting in case 2) and then gauging the $1$-form symmetry, localizing us to a given universe. 
    Indeed, the only operators we have on the boundary in this case come from taking $V(\Sigma_2)$ with the surface $\Sigma_2$ parallel to the boundary.  These operators generate the $(-1)$-form symmetry that appears after gauging the $1$-form symmetry of case 2. Inserting such an operator shifts universes in a region of the boundary, with the boundary piece $\exp(i\oint_{\partial\Sigma_2}A_1)$ of the operator acting as a domain wall, as depicted in Fig.~\ref{fig:case_3}.  

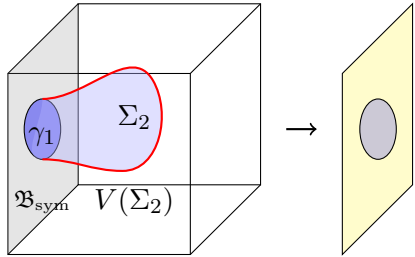
\begin{figure}[ht]
\centering
\begin{tikzpicture}[scale = 0.8]
\draw [black] (0,0,0) -- (3,0,0) -- (3,3,0) -- (0,3,0) -- cycle; 
\draw [black] (0,0,3) -- (3,0,3) -- (3,3,3) -- (0,3,3) -- cycle; 
\draw [black] (0,0,0) -- (0,0,3);
\draw [black] (3,0,0) -- (3,0,3);
\draw [black] (3,3,0) -- (3,3,3);
\draw [black] (0,3,0) -- (0,3,3);

\fill[gray, opacity=0.2] (-0.01,0,0) -- (-0.01,3,0) -- (-0.01,3,3) -- (-0.01,0,3) -- cycle; 

\fill[blue, opacity=0.4] (-0.01,1.5,1.5) ellipse (0.3 and 0.5); 

\draw [black] (-0.01,1.5,1.5) ellipse (0.3 and 0.5);  

\fill[blue!30, opacity=0.4] 
    (-0.01,2,1.5) 
    to[out=0, in=120] (2,2.5,2) 
    to[out=-60, in=0] (1.5,1,2) 
    to[out=180, in=0] (-0.01,1,1.5) 
    to[out=180, in=-90] (-0.01,2,1.5); 
    
\draw[thick, red] (-0.01,2,1.5) 
    to[out=0, in=120] (2,2.5,2) 
    to[out=-60, in=0] (1.5,1,2) 
    to[out=180, in=0] (-0.01,1,1.5);
    
\draw[thick, ->] (4,1.5,1.5) -- (4.5,1.5,1.5); 

\fill[yellow, opacity=0.2] (5.5,0,0) -- (5.5,3,0) -- (5.5,3,3) -- (5.5,0,3) -- cycle; 

\draw [black] (5.5,0,0) -- (5.5,3,0) -- (5.5,3,3) -- (5.5,0,3) -- cycle; 

\fill[blue, opacity=0.2] (5.51,1.5,1.5) ellipse (0.3 and 0.5);
\draw [black] (5.51,1.5,1.5) ellipse (0.3 and 0.5);  

\node at (-0.01,1.7,1.5) [below] {$\gamma_1$};  
\node at (-0.01, 0.3, 1.5) {\footnotesize{$\mathfrak{B}_\text{sym}$}};
\node at (1.5, 1.7, 1.5) {$\Sigma_2$};
\node at (1.5, 0.3, 1.5) {$V(\Sigma_2)$};
\end{tikzpicture}
    \caption{The figure depicts the action of $(-1)$-form symmetry, as in case 3. The operator $V(\gamma_1 =\partial \Sigma_2, \Sigma_2)$ remains parallel to the symmetry boundary due to the chosen boundary condition. Insertion of the defect $V$ in the bulk changes universe (value of the point operator) in the blue colored region.}
    \label{fig:case_3}
\end{figure}
\end{itemize}
\begin{table}[H]
    \renewcommand{\arraystretch}{1.3}
    \centering
    \begin{tabular}{|c|c|c|c|}
        \hline
        Case & Dirichlet B.C. & {Neumann B.C.} & {Symmetry} \\
        \hline
        1) & $A_1,C_2$ & $\hat{A}_1, B_0$  & Trivially Acting $\Z^{(0)}_2$ \\
        \hline
        2) & $\hat{A}_1, C_2$ & $A_1, B_0$  & $\Z^{(1)}_2$, $\hat{\Z}^{(0)}_2$\\
        \hline
        3) & $\hat{A}_1, B_0$ & $A_1,C_2$ & $\Z^{(-1)}_2$ \\
        \hline
    \end{tabular}
    \caption{Summary of the boundary conditions and the resulting symmetries in the three different cases.}
    \label{tab:summaryoftas}
\end{table}

\section{Non-Invertible \texorpdfstring{\boldmath{$(-1)$}}{(-1)}-form symmetry from Dimensional Reduction}
\label{sec:non-invertible}

So far we have only considered group-like symmetries.
As a consequence, viewing a $(-1)$-form symmetry transformation as stacking the QFT with a topological sector (the worldvolume of the spacetime-filling symmetry defect) is a bit inflated since this sector is trivial.
For a non-invertible symmetry,\footnote{We focus here on non-invertible $(-1)$-form symmetries; non-invertible $(d-1)$-form symmetries, their gauging, and their meaning for decomposition have been previously discussed in \cite{Nguyen:2021naa,Santilli:2024dyz}.} this is no longer true, and the non-trivial defect-TFT is necessary to obtain a consistent theory after shifting the ``coupling parameter'' associated with the $(-1)$-form symmetry.
In this section we study an example of non-invertible $(-1)$-form symmetry in two dimensions, where the TFT-sector is needed to trace the symmetry operators of a $0$-form $\mathbb{Z}_N$ symmetry where $N$ depends on the parameter.
We first describe the corresponding topological operators in the 3d SymTFT, and then discuss the physical interpretation from the 2d boundary theory's perspective.

\subsection{Dimensional reduction of non-invertible \texorpdfstring{$\mathbb{Q}/\mathbb{Z}$}{Q/Z} flavor symmetry}

The SymTFT is given by the bulk action
\begin{align}\label{eq:SymTFT_action_3d-ABJ}
    S_\text{bulk}^\text{3d} = \frac{1}{2\pi} \int \left( b_1 \,d A_1 + f_2 \, dB_0 + f_0 \, d B_2 + \frac{k_A}{2\pi} \, f_0 \, A_1 \, f_2 \right) \, ,
\end{align}
where $b_1, f_i$ are $\mathbb{R}$ gauge fields, $A_1, B_j$ are $U(1)$ gauge fields, and $k_A \in \mathbb{Z}$ an integer constant. Before discussing the various boundary condition choices, let us put this setting into a broader context.

One way to motivate the 3d system \eqref{eq:SymTFT_action_3d-ABJ} is to view it as the $S^2$ reduction of a 5d SymTFT for a 4d $U(1)$ gauge theory with a $U(1)^{(0)}_F$ flavor symmetry,
\begin{align}\label{5DSYM}
    S_\text{bulk}^\text{5d} = \frac{1}{2\pi} \int_{{\cal W}_5} \left(b_3dA^\text{4d}_1 + f^\text{4d}_2dB^\text{4d}_2 + \frac{k_{A}}{4\pi}A^\text{4d}_1(f^\text{4d}_2)^2 + \frac{k_{2}}{4\pi} A^\text{4d}_1dA^\text{4d}_1f^{4d}_2 + \frac{k_T}{12\pi} A^\text{4d}_1(dA^\text{4d}_1)^2\right).
\end{align}
Here, $b_3$ and $f_2$ are $\R$-valued, while $A_1$ and $B_2$ are $U(1)$-valued gauge fields,\footnote{In fact, because of the anomaly terms, it is slightly misleading to call $B_2$ an ordinary $U(1)$-valued gauge field; its transition functions will mix with the other fields in nontrivial ways.} with the following gauge transformations:
\begin{align}\label{eq:gauge-transformations-5dSYMTFT}
\begin{split}
    A^\text{4d}_1\ \longrightarrow\ & A^\text{4d}_1+d\lambda_0\, ,\\
    B^\text{4d}_2\ \longrightarrow\ & B^\text{4d}_2+d\lambda_1+\frac{k_A}{2\pi}A^\text{4d}_1\mu_1-\frac{k_{A}}{2\pi}f^\text{4d}_2\lambda_0+\frac{k_{A}}{2\pi}d\lambda_0\mu_1 \, ,\\
    f^\text{4d}_2\ \longrightarrow\ & f^\text{4d}_2+d\mu_1 \, ,\\
    b_3\ \longrightarrow\ & b_3+d\mu_2-\frac{k_{A}}{2\pi}f^\text{4d}_2\mu_1-\frac{k_{2}}{4\pi}dA^\text{4d}_1\mu_1-\frac{k_{2}}{4\pi}f^\text{4d}_2d\lambda_0-\frac{k_T}{12\pi}dA^\text{4d}_1d\lambda_0-\frac{k_{A}}{4\pi}\mu_1d\mu_1\, .
\end{split}
\end{align}
The $U(1)$ gauge symmetry is reflected having Dirichlet/Neumann boundary conditions for $B_2^\text{4d} / f_2$ on the symmetry boundary \cite{Antinucci:2024zjp}, and the $U(1)_F^{(0)}$ flavor symmetry has background field $A_1^\text{4d}$ which has Dirichlet boundary conditions.
We will focus on the ABJ anomaly term with coefficient $k_A$, and shall turn off possible 2-group and self-anomaly of the flavor symmetry by setting $k_2 = k_T = 0$.
In the 4d QFT, a non-zero integer $k_A$ modifies the $U(1)$ flavor symmetry into a non-invertible $\mathbb{Q}/\mathbb{Z}$ \cite{Choi:2022jqy, Cordova:2022ieu}.

We are interested in placing the theory on ${\cal W}_5 = {\cal W}_3 \times S^2$ with the reduction ansatz
\begin{align}
    b_3 = b_1 \wedge \text{vol}_{S^2} \, , \quad f_2^\text{4d} = f_0 \wedge \text{vol}_{S^2} + f_2 \, , \quad B_2^\text{4d} = B_2 + B_0 \wedge \text{vol}_{S^2} \, , \quad A_1^\text{4d} = A_1 \, ,
\end{align}
which colloquially is also written as,
\begin{align}
    \int_{S^2}b_3=b_1,\qquad\int_{S^2} f^\text{4d}_2 = f_0,\qquad\int_{S^2} B^\text{4d}_2=B_0 \, , \qquad \int_{S^2} dA_1 = 0 \, .
\end{align}
Such a reduction has been previously considered in \cite{Nardoni:2024sos}, where the authors set $f_0$, which takes values $2\pi m_G \in 2\pi \mathbb{Z}$, to be constant on ${\cal W}_3$.
The resulting SymTFT is therefore that of a sector with fixed gauge-flux on $S^2$ which is a universe associated with the 2d 1-form symmetry generated by $\exp(i \beta f_0)$ \cite{Nardoni:2024sos}.
In order to describe the transitions between such sectors as a $(-1)$-form symmetry action, we also allow for $f_0$ to fluctuate in ${\cal W}_3$, which leads to the 3d SymTFT we started with.

Irrespective of (but corroborated by) this higher-dimensional origin, we therefore view \eqref{eq:SymTFT_action_3d-ABJ} as the SymTFT of a 2d $U(1)$ gauge theory (with $f_2$ the gauge field) and a 0-form global symmetry with background field $A_1$, which have an ABJ-like anomaly $A_1 f_2$ whose coefficient $k_A f_0$ varies with an $\mathbb{R}$-valued scalar field $f_0$ which is quantized to be $2 \pi \mathbb{Z}$ valued.
In 2d this kind of anomaly will only leave an invertible $\mathbb{Z}_{k_A f_0}$ in a fixed-flux sector \cite{Nardoni:2024sos}, but we will show that instead the $(-1)$-form symmetry shifting between these sectors is non-invertible.
The absolute 2d theory which will exhibit this non-invertible symmetry is realized with Dirichlet boundary conditions for $A_1, B_0, f_0$.

\subsection{Non-Invertible \texorpdfstring{$(-1)$}{(-1)}-form Symmetry}

\subsubsection*{Topological Operators in the SymTFT bulk}

We begin by constructing the topological operators of the SymTFT \eqref{eq:SymTFT_action_3d-ABJ}.
For that, we need the gauge transformations of the various fields, which can also be derived from the gauge transformations of the 5d SymTFT (using \eqref{eq:gauge-transformations-5dSYMTFT} with $k_T = k_2 =0$, $\mu_0=\int_{S^2}\mu_2$, and $n=\frac{1}{2\pi}\int_{S^2}d\lambda_1\in\Z$).
These are
\begin{equation}\label{eq:gauge-trafo-3d} 
\begin{aligned}
    f_0\ \longrightarrow\ & f_0 \, ,\\
    A_1\ \longrightarrow\ & A_1+d\lambda_0 \, ,\\
    f_2\ \longrightarrow\ & f_2+d\mu_1 \, ,\\
    B_0\ \longrightarrow\ & B_0+2\pi n-\frac{k_{A}}{2\pi}f_0\lambda_0 \, ,\\
    b_1\ \longrightarrow\ & b_1+d\mu_0-\frac{k_{A}}{2\pi}f_0\mu_1 \, ,\\
    B_2\ \longrightarrow\ & B_2+d\lambda_1+\frac{k_{A}}{2\pi}A_1\mu_1-\frac{k_{A}}{2\pi}f_2\lambda_0+\frac{k_{A}}{2\pi}d\lambda_0\mu_1 \, .
\end{aligned}
\end{equation}
The equations of motion read as,
\begin{align}\label{eq:eom_3d_SymTFT}
    \begin{split}
        d A_1 = d f_0 = d f_2 & = 0 \, , \\
        d B_0 + \frac{k_A}{2\pi} \, f_0 \, A_1 & = 0 \, , \\
        d B_2 + \frac{k_A}{2\pi} \, A_1 \, f_2  & = 0 \, , \\
        d b_1 + \frac{k_A}{2\pi} \, f_0 \, f_2 & = 0 \, .
    \end{split}
\end{align}
Using the gauge transformations and equations of motion, we can write down gauge-invariant topological operators in the bulk of the SymTFT.  The fields $f_0$, $A_1$, and $f_2$ only transform conventionally and give rise to the usual operators,
\begin{equation}
    \mathfrak{V}_\alpha[\Sigma_2]=e^{i\alpha\oint_{\Sigma_2}f_2} \, ,\qquad X_n[\gamma_1]=e^{in\oint_{\gamma_1}A_1} \, ,\qquad {U}_\beta[x]=e^{i\beta f_0(x)} \, .
\end{equation}
Here $n\in\Z$ in order to ensure invariance under large gauge transformations of $A_1$, and $\alpha,\beta\in\R/\Z$ since the sum over the gauge bundle degrees of freedom of $B_0$ and $B_2$ force $f_0$ and the periods of $f_2$ to be valued in $2\pi\Z$.

The other three fields require more subtlety to build corresponding topological operators.  The results are qualitatively different depending on whether $k_{A}$ is zero or nonzero.  If $k_{A}=0$, then we have three more genuine operators,
\begin{equation}\label{eq:non-invariant-ops}
    \mathfrak{U}_m[x]=e^{imB_0(x)} \, , \qquad W_\alpha[\gamma_1]=e^{i\alpha\oint_{\gamma_1} b_1} \, , \qquad {Z}_p[\Sigma_2]=e^{ip\oint_{\Sigma_2}B_2} \, .
\end{equation}
Here $m$ and $p$ are integers, so that the corresponding operators are well-defined under large gauge transformations, while $\alpha$ is valued in $\R/\Z$ since the sum over the gauge bundle degrees of freedom of $A_1$ enforces that $\oint b_1\in 2\pi\Z$.

When $k_{A}\ne 0$, these last operators are not gauge invariant.  In this case we can build gauge-invariant versions by making the operators non-genuine,
\begin{align}\label{eq:gauge-invariant-attachments}
\begin{split}
    \mathfrak{U}_m[\{x\}\cup\overline{\{y\}}=\sigma_0=\partial\mathcal{C}_1;\mathcal{C}_1]=\ & e^{im\left(B_0(x)-B_0(y)\right)+im\frac{k_{A}}{2\pi}\int_{\mathcal{C}_1}f_0A_1} \, ,\\
    W_\alpha[\gamma_1=\partial\mathcal{S}_2;\mathcal{S}_2]=\ & e^{i\alpha\oint_{\gamma_1}b_1+i\alpha\frac{k_{A}}{2\pi}\int_{\mathcal{S}_2}f_0f_2} \,,\\
    {Z}_p[\Sigma_2=\partial\mathcal{M}_3;\mathcal{M}_3]=\ & e^{ip\oint_{\Sigma_2}B_2+ip\frac{k_{A}}{2\pi}\int_{\mathcal{M}_3}A_1f_2}\, .
\end{split}
\end{align}
These are both topological and gauge invariant if we use the equations of motion (in particular, $df_0=0$ and $df_2=0$ are required for gauge-invariance).
Once again $m$ and $p$ are integers.
Note also that if $\alpha k_{A}\in\Z$, or $\alpha k_{A}m_G\in\Z$ in a fixed-flux sector with $f_0=2\pi m_G$, then $W_\alpha[\gamma_1;\mathcal{S}_2]$ does not actually depend on a choice of bounding surface $\mathcal{S}_2$, since two different choices will differ by $\exp[i\alpha k_{A}m_G\oint_{\mathcal{S}_2^{(1)}\cup\overline{\mathcal{S}_2^{(2)}}}f_2]=1$, using the quantization of periods of $f_2$.  A downside of this construction is that these non-genuine operators (even $W_\alpha$ for $\alpha\in\frac{1}{k_{A}}\Z$) require a bounding manifold ${\cal S}_2$ to be well-defined.

To understand the physical consequences of these operators, we first note that $f_0$ appears as a background field from the perspective of these non-genuine operators, so we investigate their properties in a region of the 3d bulk where $f_0 = 2\pi m_G$, $m_G \in \mathbb{Z}$ is constant --- after all $f_0$ is locally constant by the equations of motion.
If this region is the entirety of ${\cal W}_3$, then we have the same setting as in \cite{Nardoni:2024sos}, namely the SymTFT for a fixed-flux sector.
In either case, the effect of the $\mathfrak{U}_m$ operators is to enforce that $X_n[\gamma_1]$ is a trivial operator whenever $n$ is a multiple of $k_{A}m_G$, since at those values the operator can be opened up by nucleating a pair of $B_0$ insertions at its end-points, thus the operator cannot link non-trivially with anything.  Similarly, the effect of $W_\alpha$ is that $\mathfrak{V}_\alpha[\Sigma_2]$ is trivial (cannot link with anything) for any value of $\alpha$.
That is, there is no ``magnetic'' $(-1)$-form symmetry generated by these operators.

Finally, what goes beyond the discussion of a fixed-flux sector is the inclusion of the operators $Z_p$.
The operator tells us that we can add a term $ip\frac{k_{A}}{2\pi}\int A_1f_2$ to the bulk action in a region that is surrounded by $\Sigma_2$.  This effectively shifts the value of $f_0$ by $2\pi p$ in the interior region.
In this way the SymTFT \eqref{eq:SymTFT_action_3d-ABJ} incorporates transitions between flux sectors.

To obtain operators that eventually generate symmetries in the absolute theory, we must find genuine versions of these operators:
starting from the operators \eqref{eq:non-invariant-ops} which are \emph{not} gauge-invariant for $k_A \neq 0$, we can recover invariance by stacking local degrees of freedom. 
This would also allow us to define the operators on any manifold regardless of the existence of suitable higher-dimensional attachments such as in \eqref{eq:gauge-invariant-attachments}.  

For example, we can make a gauge-invariant version of $\mathfrak{U}_m$ by introducing a local $U(1)$-valued field $\phi_0$ whose gauge transformtion is $\phi_0\rightarrow\varphi_0+2\pi q+\lambda_0$, where $q\in\Z$ and $\lambda_0$ is the parameter for gauge transformations of $A_1$, see \eqref{eq:gauge-trafo-3d}.  Then we can define (not being careful about normalization)
\begin{equation}
    \widetilde{\mathfrak{U}}_m[x]=\int\mathcal{D}\phi_0e^{im\left(B_0(x)+\frac{k_{A}}{2\pi}f_0(x)\phi_0\right)}.
\end{equation}
This operator is now gauge-invariant by construction, but it does not give us anything profound; the path integral over $\phi_0$, which is really just an ordinary integral from $0$ to $2\pi$, simply produces a Dirac delta,
\begin{equation}\label{gen_p_op}
    \widetilde{\mathfrak{U}}_m[x]=\delta_{f_0,0}e^{imB_0(x)}.
\end{equation}
One way to view the role of $\phi_0$ is to construct a projection operator onto the $f_0=0$ sector.
Another, more convenient interpretation is that the point operators $\widetilde{\mathfrak{U}}_m$ are trivial if $f_0 \neq 0$, related to the observation above that the surface operators $\mathfrak{V}_\alpha = e^{i\alpha \oint f_2}$ act trivially, i.e., have no non-trivial point operators which they link with.

A similar construction applies to $W_\alpha$.  We introduce a new $U(1)$ field $\phi_1$ with transformations $\phi_1\rightarrow\phi_1+d\chi_0+\mu_1$, with $\mu_1$ being the gauge parameter of $f_2$ in \eqref{eq:gauge-trafo-3d}.  Then we can write down
\begin{equation}
    \widetilde{W}_\alpha[\gamma_1]=\int\mathcal{D}\phi_1e^{i\alpha\oint_{\gamma_1}\left(b_1+\frac{k_{A}}{2\pi}f_0\phi_1\right)}.
\end{equation}
Again, the path integral over $\phi_1$ simply produces a Dirac delta which forces $f_0=0$, i.e., a projection operator onto this sector,
\begin{equation}
    \widetilde{W}_\alpha[\gamma_1]=\delta_{f_0,0}e^{i\alpha\oint_{\gamma_1}b_1} \, . \label{eq:0-form_generator_f0=0}
\end{equation}
Equivalently, these operators are the generators of the $U(1)$-flavor symmetry which only persists in the $f_0 =0$ sector where the ABJ anomaly is absent.

Finally, playing the same trick with $Z_p$ leads to
\begin{equation}\label{eq:codim-1-genuine}
    {V}_p[\Sigma_2]=\int\mathcal{D}\phi_0\mathcal{D}\phi_1e^{ip\oint_{\Sigma_2}\left(B_2+\frac{k_{A}}{2\pi}\left(f_2\phi_0-A_1\phi_1-\phi_0d\phi_1\right)\right)}.
\end{equation}
Here the local fields $\phi_0$ and $\phi_1$ have the same gauge transformations as above, i.e.
\begin{equation*}
    \begin{aligned}
        \phi_0 &\rightarrow\phi_0+2\pi q+\lambda_0 \, , \\
        \phi_1 &\rightarrow\phi_1+d\chi_0+\mu_1 \, ,
    \end{aligned}
\end{equation*}
where $\lambda_0$ and $\mu_1$ are the bulk gauge parameters in  \eqref{eq:gauge-trafo-3d}.
In this case the fields $\phi_{0,1}$ are not simply Lagrange multipliers, but are dynamical fields on the worldsheet of the defect.
They give rise to a 2d BF-theory, whose 0- and 1-form global symmetries couple to the bulk fields $A_1$ and $f_2$, respectively, as background fields \cite{Antinucci:2024zjp}.

Since $\phi_0$ and $\phi_1$ transform under the bulk gauge transformations, the corresponding operators $u_n[x]=e^{in\phi_0(x)}$ and $v_m[\gamma_1]=e^{im\oint_{\gamma_1}\phi_1}$ that can appear on the defect must necessarily attach to bulk operators, i.e., a bulk $X_n$ line operator can terminate on the defect on a $u_n[x]$ point operator, while the $v_m[\gamma_1]$ lines must sit at the boundary of a bulk surface operator involving $f_2$.  To be specific, we can either have an operator $\mathfrak{V}_m[\mathcal{S}_2]$ with $\partial\mathcal{S}_2=\gamma_1$ (such an operator would be trivial if it had no boundary), or in a region where $f_0=2\pi m_G\ne 0$ we can have a $W_{\alpha}[\gamma_1',\mathcal{S}_2]$ with $\alpha k_A m_G \in \mathbb{Z}$, where $\mathcal{S}_2$ stretches between $\gamma_1$ on the defect and $\gamma_1'$ in the bulk.

To summarize, starting from the SymTFT action \eqref{eq:SymTFT_action_3d-ABJ}, which we rewrite again for convenience, 
\begin{align}\label{3DNONINV}
    S_\text{bulk}^\text{3d} = \frac{1}{2\pi} \int \left( b_1 \,d A_1 + f_2 \, dB_0 + f_0 \, d B_2 + \frac{k_A}{2\pi} \, f_0 \, A_1 \, f_2 \right) \, ,
\end{align}
The gauge invariant topological operators are,
\begin{equation}\label{3dbulkop}
    \begin{aligned}
        \mathfrak{V}_\alpha [\Sigma_2] &= e^{i \alpha \oint_{\Sigma_2} f_2} \, , \\
        X_n [\gamma_1]  &= e^{i n \oint_{\gamma_1} A_1} \, , \quad {U}_\beta [x]  = e^{i \beta \, f_0(x)} , \\
        \mathfrak{U}_m [y,\ell_1] &= \exp(i m \big( B_0(y) + \textstyle\frac{k_A}{2\pi} \int_{\ell_1} f_0 A_1 \big)) \, , \quad \partial \ell_1 = y \, , \\
        W_\alpha [\gamma_1, {\cal S}_2] & = \exp(i \alpha \big( \textstyle\oint_{\gamma_1} b_1 + \frac{k_A}{2\pi} \int_{{\cal S}_2} f_0 f_2 \big)) \, , \quad \partial{\cal S}_2 = \gamma_1 \, ,\\ 
        {Z}_p [\Sigma_2, M_3] & = \exp( i p \big( \textstyle\oint_{\Sigma_2} B_2 + \frac{k_A}{2\pi} \int_{M_3} A_1 f_2 \big) ) \, , \quad \partial M_3 = \Sigma_2 \, ,
    \end{aligned}
\end{equation}
where $\alpha, \beta \in \mathbb{R}/\mathbb{Z}$ and $m,n,p \in \mathbb{Z}$.
We have found genuine counterparts $\widetilde{\mathfrak{U}}, \widetilde{W}, V$ for the non-genuine operators $\mathfrak{U}, W, Z$, where the first two are only non-trivial for $f_0 = 0$.
The third operator given by \eqref{eq:codim-1-genuine} is the codimension-1 defect one that shifts $f_0$.
The topological sector localized on its worldvolume contains point operators $u_n$ on which the bulk $X_n$ lines can end, and line operators $v_m[\gamma_1]$ on the bulk operators $W_\alpha[\gamma_1', {\cal S}_2]$ can end, provided $\partial {\cal S}_2 = \gamma_1' \cup \gamma_1$ and $\alpha k_A \frac{f_0}{ 2 \pi} = m \in \mathbb{Z}$.

\subsubsection*{Non-invertible codimension-1 defects}

Next we demonstrate that the genuine codimension-1 operator \eqref{eq:codim-1-genuine} is non-invertible.
This is due to the non-trivial TQFT on its worldvolume which is a 2d $\mathbb{Z}_{p k_A}$-BF-theory $\mathcal{A}$,
\begin{align}\label{eq:worldvolume_Vp-action}
\begin{split}
    {V}_p[\Sigma_2] & = \int {\cal D} \phi_0 {\cal D} \phi_1 \exp[ i p  \textstyle\oint_{\Sigma_2} \big( B_2 + \tfrac{k_A}{2\pi}( \phi_0 f_2 -  A_1\phi_1 - \phi_1 d \phi_0 \big)] \\
    & \equiv \exp(i p \textstyle\oint_{\Sigma_2} B_2) \, \mathcal{A}[\Sigma_2; A_1, f_2] \, .
\end{split}
\end{align}
As emphasized earlier, since $\phi_0$ and $\phi_1$ transforms under bulk gauge transformations, the point operator $e^{in\phi_0}$ must be attached to a line. Therefore, there are no genuine topological point operators living on the defect ensuring that this is in fact a simple codimension-1 defect.

\paragraph{Fusion:}
In order to establish the non-invertible nature of this defect we would like to compute the fusion, 
\begin{equation}
    \begin{aligned}
V_m \otimes V_n &\cong \int \exp(i(m+n)\int  B_2 + \frac{ik_Am}{2\pi}(f_2 \phi_0 -A_1\phi_1 -\phi_1d\phi_0) + \frac{ik_An}{2\pi}(f_2 \phi'_0 -A_1\phi'_1 -\phi'_1d\phi'_0) ) \\
&\cong \int \exp\int (iA + \frac{ik_A}{2\pi}(mf_2 \phi_0 + nf_2\phi'_0 - mA_1\phi_1 - nA_1\phi'_1 - m\phi_1d\phi_0 -n\phi'_1d\phi'_0) ) \\
&\cong \int \exp\int (iA + \frac{ik_A}{2\pi} ((m\phi_0 + n\phi'_0)f_2 - (m\phi_1 + n\phi'_1)A_1 - m\phi_1d\phi_0 -n\phi'_1d\phi'_0 ) )
    \end{aligned}
\end{equation}
where $A = (m+n) B_2$. We do the following field redefinitions, 
\begin{equation}
    \begin{aligned}
        \phi_0 &= \bar{\phi}_0, \quad \check{\phi}_0 = \phi'_0 - \phi_0 \\
        \phi_1 &= \bar{\phi}_1, \quad \check{\phi}_1 = \phi'_1 -\phi_1
    \end{aligned}
\end{equation}
working with each term at a time in the exponential, 
\begin{equation*}
    \begin{aligned}
        (m\phi_0 + n\phi'_0)f_2 &= (m\bar{\phi}_0 + n\bar{\phi}_0 + n\check{\phi}_0)f_2 = (m+n)\bar{\phi}_0f_2 + n\check{\phi}_0f_2 \\
        (m\phi_1 + n\phi'_1)A_1 &= (m\bar{\phi}_1 + n\bar{\phi}_1 + n\check{\phi}_1) A_1 = (m+n)\bar{\phi}_1A_1 + n\check{\phi}_1 A_1 \\
        m\phi_1d\phi_1 &= m\bar{\phi}_1 d\bar{\phi}_0 \\
        n\phi'_1d\phi'_0 &= n(\check{\phi}_1 + \bar{\phi}_1)d(\check{\phi}_0 + \bar{\phi}_0)
    \end{aligned}
\end{equation*}
Collecting all the terms together and using $A = (m+n) B_2$, 
\begin{equation}
    \begin{aligned}
       V_m \otimes V_n \cong  \int \exp \int \underbrace{(iA + \frac{ik_A}{2\pi}((m+n)\bar{\phi}_0f_2 - (m + n)\bar{\phi}_1A_1) -(m+n) \bar{\phi}_1d\bar{\phi}_0 )}_{V_{m+n}} \times \\
       \exp \int \frac{ink_A}{2\pi}(\check{\phi}_0f_2 - \check{\phi}_1A_1 - \check{\phi}_1d\check{\phi}_0 - \check{\phi}_1d\bar{\phi}_0 -\bar{\phi}_1 d\check{\phi}_0)
    \end{aligned}
\end{equation}

Specializing to $n=-m$ will produce the non-invertible fusion,  
\begin{equation}
    \begin{aligned}
         V_m \otimes V_{-m} \cong  \int \exp \int \frac{-imk_A}{2\pi}(\check{\phi}_0f_2 - \check{\phi}_1A_1 - \check{\phi}_1d\check{\phi}_0 - \check{\phi}_1d\bar{\phi}_0 -\bar{\phi}_1 d\check{\phi}_0)
    \end{aligned}
\end{equation}
Performing the path integrals over $\bar{\phi}_0$ and $\bar{\phi}_1$ ensure that $\check{\phi}_1$ and $\check{\phi}_0$ are $\Z_{mk_A}$-valued fields, so the right-hand side of this equation is just the partition function of some $\Z_{mk_A}$ gauge theory coupled to the bulk fields $A_1$ and $f_2$.

Since $\check{\phi}_0$ is gauge invariant, we have genuine point operators $\check{u}_n=e^{in\check{\phi}_0}$ which live on the defect.  Because of the quantization of $\check{\phi}_0$, $\check{u}_n\sim\check{u}_{n+k_Am}$, so there are $k_Am$ distinct point operators, and hence the theory on this defect decomposes into $k_Am$ different universes.  This phenomenon was previously discussed in the context of fusion of codimension-1 condensation defects in~\cite{Lin:2022xod}.

On a single defect $V_m$ we had point operators $u_n$ which had to attach to $X_n$ lines in the bulk.  In the fusion between $V_m$ and $V_{-m}$, we can have the $X_n$ line stretching from one defect to the other; when we fuse the defects together, this gives rise to the genuine point operators and hence decomposition.

\subsection{Changing 0-form symmetry across universes}

With the operators $V_p$, we now have a genuine codimension-1 defect across which the locally constant value of $f_0$ changes by $2\pi p$.
Using it, we aim to understand how the 0-form symmetry is realized in each universe (i.e., gauge flux sector from the dimensional reduction perspective) and, more importantly, how it is modified within the SymTFT description as we shift between the universes.

First, let us recall the boundary condition choices of Dirichlet for $A_1, B_0, f_0$, giving us a $U(1)$ gauge field with dynamical field strength $f_2$ and a 0-form global symmetry with background field $A_1$ in the 2d absolute theory.
The corresponding symmetry generators are associated with $e^{i \alpha \oint b_1}$; indeed we have seen that for $f_0 = 0$, these line operators are gauge invariant for any $\alpha \in U(1)$.

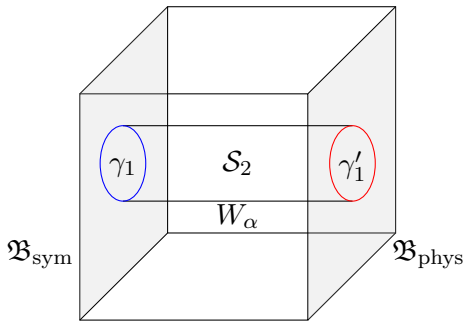
\begin{figure}[ht]
    \centering
    \begin{tikzpicture}[scale = 1.0]

    \draw [black] (0,0,0) -- (3,0,0) -- (3,3,0) -- (0,3,0) -- cycle; 
    \draw [black] (0,0,3) -- (3,0,3) -- (3,3,3) -- (0,3,3) -- cycle; 
    \draw [black] (0,0,0) -- (0,0,3);
    \draw [black] (3,0,0) -- (3,0,3);
    \draw [black] (3,3,0) -- (3,3,3);
    \draw [black] (0,3,0) -- (0,3,3);

    \fill[lightgray, opacity=0.2] (-0.01,0,0) -- (-0.01,3,0) -- (-0.01,3,3) -- (-0.01,0,3) -- cycle; 
    \fill[lightgray, opacity=0.2] (3.01,0,0) -- (3.01,3,0) -- (3.01,3,3) -- (3.01,0,3) -- cycle; 

    \draw [black] (-0.01,2,1.5) -- (3.01,2,1.5);
    \draw [black] (3.01,1,1.5) -- (-0.01,1,1.5);
    \draw [blue] (-0.01,1.5,1.5) ellipse (0.3 and 0.5);  
    \draw [red] (3.01,1.5,1.5) ellipse (0.3 and 0.5);   

    \node at (-0.01,1.7,1.5) [below] {$\gamma_1$};  
    \node at (3.01,1.8,1.5) [below] {$\gamma_1'$};  

    \node at (1.5,1.8,1.5) [below] {${\mathcal{S}}_2$};  
    \node at (1.5, 0.8, 1.5) {$W_\alpha$};
    \node at (-1.1, 0.3, 1.5) {$\mathfrak{B}_\text{sym}$};
    \node at (4.01, 0.3, 1.5) {$\mathfrak{B}_\text{phys}$};
\end{tikzpicture}
    \caption{Realization of topological lines $W_\alpha[\gamma_1]$ generating the 0-form symmetry for $f_0 = m_G$:
    The cylinder ${\cal S}_2$ with flux $\frac{\alpha k_A m_G}{2\pi} \int f_2$ terminates on the (blue) line operator $\exp(i \alpha \oint_{\gamma_1} b_1)$ which can be pushed to the symmetry boundary due to Neumann boundary conditions, and is anchored on the physical boundary on the (red) Wilson line $\exp(i \frac{\alpha k_A m_G}{2\pi} \oint_{\gamma_1'} a)$.
    This tube operator is topological and gauge-invariant provided $\alpha k_A m_G \in \Z$, and collapses upon interval compactification to a line operator on $\gamma_1 \simeq \gamma_1'$.}
    \label{fig:0-form-Sym}
\end{figure}

In a local bulk region with constant $f_0 \neq 0$, we only have the non-genuine operators 
\begin{equation}
    W_\alpha [\gamma_1, {\cal S}_2]  = \exp(i \alpha \big( \textstyle\oint_{\gamma_1} b_1 + \frac{k_A f_0}{2\pi} \int_{{\cal S}_2} f_2 \big))
\end{equation}
at our disposal.
However, for $\alpha$ such that $\alpha k_A f_0 \in \mathbb{Z}$ the dependence on the bounding surface ${\cal S}_2$ drops out, so that the endpoints of such $W_\alpha[\gamma_1 = \partial {\cal S}_2] \equiv W_\alpha[\gamma_1]$ give a genuine topological line operator \cite{Kapustin:2014gua}.
When $\gamma_1$ is a non-contractible 1-cycle on the boundary, i.e., when there is no 2-chain ${\cal S}_2 \subset \mathfrak{B}_\text{sym}$ with $\partial {\cal S}_2 = \gamma_1$, the construction of this genuine line operator requires more care.
Assuming for now that $f_0 \equiv 2\pi m_G \neq 0$ is constant along the entire bulk ${\cal W}_3 \cong \mathfrak{B} \times [0,1]$ (so it must be also the Dirichlet boundary value of $f_0$), we can still have a $W_\alpha[\gamma_1]$ operator wrapping that line $\gamma_1 \subset \mathfrak{B} \times \{0\}$ if we attach it to a cylinder $\mathcal{S}_2$ that stretches all the way to the physical boundary and ends on $\gamma_1 \simeq \gamma_1' \subset \mathfrak{B} \times \{1\}$, as depicted in Fig.~\ref{fig:0-form-Sym}.
The end-locus $\gamma_1' \subset \mathfrak{B}_\text{phys}$ is wrapped by a Wilson line $\exp(i \frac{\alpha k_A m_G}{2\pi} \oint_{\gamma_1'} a)$ with $a$ the $U(1)$ gauge connection on the physical boundary, for which the quantization condition $\alpha k_A m_G \in \Z$ now ensures that it is invariant under large gauge transformations of $f_2 = da$.
As long as $\alpha k_A m_G\in\Z$, the entire ``tube'' operator is insensitive to the choice of the cylinder ${\cal S}_2$.
After collapsing the interval, where ${\cal S}_2$ is squashed so that the Wilson line and the $b_1$-holonomy sit on top of each other, this construction realizes $\mathbb{Z}_{k_A m_G}$ $0$-form symmetry generators $W_\alpha[\gamma_1]$ on non-contractible 1-cycles $\gamma_1 \cong \gamma_1'$:
\begin{align}\label{eq:0-fs-gen}
    W_\alpha[\gamma_1] = \exp(i \alpha \textstyle\oint_{\gamma_1} \big(  b_1 - \frac{k_A m_G}{2\pi} a \big) ) \, , \quad \alpha k_A m_G \in \Z \, .
\end{align}
This is the same operator that was derived in \cite{Nardoni:2024sos} from the $S^2$-reduction of the 4d non-invertible $\mathbb{Q}/\Z$ symmetry generators, which yields a non-zero operator only when $\alpha k_A m_G \in \mathbb{Z}$.\footnote{This identification becomes manifest if we identify $b_1 = \star j$.
The non-conservation of this current, $d \star j = \frac{k_A m_G}{2\pi} f_2$, is due to the 2d ABJ anomaly which presently follows from the equations of motion \eqref{eq:eom_3d_SymTFT} in the SymTFT bulk.
}

The charged point operators of this 0-form symmetry arise in the conventional way, by stretching $\delta_1$ in  $X_n[\delta_1] = e^{i n \int_{\delta_1} A_1}$ from the symmetry boundary, where $A_1$ has Dirichlet boundary conditions, to the physical boundary where it ends on a suitable point operator $o_n$.
The linking phase\footnote{We assume for simplicity that the geometric linking factor is $\text{Link}(\delta_1,\gamma_1)=1$ in the exponential \eqref{eq:commutation_BF-operators}, and suppress the dependence on the cycles in the following.}
\begin{align}
    \langle X_n, W_\alpha \rangle = \exp(2\pi i n \alpha) = \exp(2 \pi i \frac{n q}{k_A m_G} ) \, , \quad q = \alpha k_A m_G \in \mathbb{Z} \, ,
\end{align}
with \eqref{eq:0-fs-gen} stems from the general non-trivial linking between $A_1$-lines and $b_1$-lines in the SymTFT.

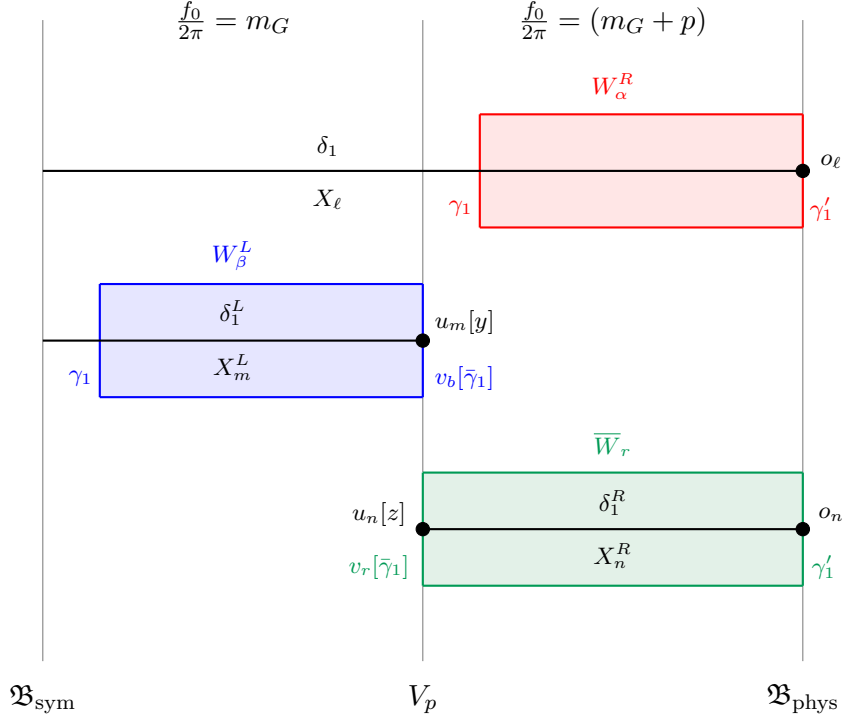
\begin{figure}[ht]
    \centering
    \begin{tikzpicture}[scale=2.5]
    \draw[gray] (-2,-.4) -- (-2,3);
    \draw[gray] (0,-.4) -- (0,3);
    \draw[gray] (2,-.4) -- (2,3);

    \node at (-1, 3) {{$\frac{f_0}{2\pi}=m_G$}};
    \node at (1, 3) {{$\frac{f_0}{2\pi}=(m_G + p)$}};

    \node at (0, -.6) {$V_p$};
    \node at (-2, -.6) {$\mathfrak{B}_{\text{sym}}$};
    \node at (2, -.6) {$\mathfrak{B}_{\text{phys}}$};

    \fill[color=blue!10] (-1.7,1) rectangle (0,1.6);
    \draw[blue, thick] (-1.7,1.6) -- (0,1.6); 
    \draw[blue, thick] (-1.7,1) -- (0,1);
    \draw[blue, thick] (-1.7, 1) -- (-1.7, 1.6); 
    \draw[thick, blue] (0,1) -- (0,1.6); 
    \node at (-1, 1.75) {\textcolor{blue}{\footnotesize{$W^L_\beta$}}};
    \node at (-1.8, 1.1) {\textcolor{blue}{\footnotesize{$\gamma_1$}}};
    \node at (0.22, 1.1) {\textcolor{blue}{\footnotesize{$v_b[\bar\gamma_1]$}}};

    \draw[thick] (-2,1.3) -- (0,1.3);
    \filldraw (0,1.3) circle (1pt);
    \node at (0.22,1.4) {\footnotesize{$u_m[y]$}};
    \node at (-1, 1.16) {\footnotesize{$X_m^L$}};
    \node at (-1, 1.45) {\footnotesize{$\delta_1^L$}};

    \fill[color=red!10] (0.3, 1.9) rectangle (2, 2.5);
    \draw[red, thick] (0.3,1.9) -- (0.3,2.5);
    \draw[red, thick] (2, 1.9) -- (2, 2.5); 
    \draw[red, thick] (0.3,1.9) -- (2,1.9); 
    \draw[red, thick] (0.3,2.5) -- (2,2.5);
    \node at (0.2, 2) {\textcolor{red}{\footnotesize{$\gamma_1$}}};
    \node at (2.1, 2) {\textcolor{red}{\footnotesize{$\gamma_1'$}}};
    \node at (1, 2.65) {\textcolor{red}{\footnotesize{$W^R_\alpha$}}};

    \draw[thick] (-2,2.2) -- (2,2.2);    
    \node at (-0.5, 2.05) {\footnotesize{$X_\ell$}};
    \node at (-0.5, 2.33) {\footnotesize{$\delta_1$}};
    \filldraw (2,2.2) circle (1pt);
    \node at (2.15,2.25) {\footnotesize{$o_\ell$}};

    \fill[color=ForestGreen!10] (0,0) rectangle (2,0.6);
    \draw[ForestGreen, thick] (0,0) -- (2,0); 
    \draw[ForestGreen, thick] (0,0.6) -- (2, 0.6);
    \draw[ForestGreen, thick] (0,0) -- (0,0.6);
    \draw[ForestGreen, thick] (2,0) -- (2,0.6);
    \node at (1, 0.75) {\color{ForestGreen}{\footnotesize{$\overline{W}_r$}}};
    \node at (- 0.23, 0.1) {\color{ForestGreen}{\footnotesize{$v_r[\bar\gamma_1]$}}};
    \node at (2.11, 0.1) {\color{ForestGreen}{\footnotesize{$\gamma_1'$}}};

    \draw[thick] (0,.3) -- (2,.3);    
    \node at (1, 0.16) {\footnotesize{$X^R_n$}};
    \node at (1, 0.44) {\footnotesize{$\delta^R_1$}};
    \filldraw (0,.3) circle (1pt);
    \node at (-0.23,.38) {\footnotesize{$u_n[z]$}};    
    \filldraw (2,.3) circle (1pt);
    \node at (2.15,.38) {\footnotesize{$o_n$}};

\end{tikzpicture}
    \caption{Configurations of various line operators $X$ and tube operators $W$ built around $e^{im \int A_1}$ and $e^{i r \int f_2}$, respectively, in the presence of a codimension-1 defect $V_p$.
    These operators rely on the point operators $u_m$ and lines $v_n[\bar\gamma_1]$ on $V_p$ on which the various bulk objects can terminate.
    On the left (symmetry-boundary-adjacent) side where $f_0 = 2\pi m_G$, we have lines $X_m^L[\delta_1^L]$ which terminates on the symmetry boundary by Dirichlet boundary conditions for $A_1$ and on a point $u_m[y]$ on $V_p$, as well as cylindrical tubes $W_\beta^L$ whose ends consists of a $b_1$-line on $\gamma_1$ in the bulk and a $v_b[\bar\gamma_1]$-line on $V_p$ and which are topological if $\beta m_G k_A \in \mathbb{Z}$; by Neumann boundary conditions the $b_1$-line can pushed into $\mathfrak{B}_\text{sym}$.
    On the right (physical-boundary-adjacent), we have lines $X_n^R$ that terminate on $u_n[z]$ on $V_p$ and $o_n$ on $\mathfrak{B}_\text{phys}$, as well as tubes $W_\alpha^R$ (topological if $\alpha (m_G + p) k_A \in \mathbb{Z}$) which are analogous to $W_\beta^L$ in that they have a free-floating $\int_{\gamma_1} b_1$-end in the bulk, but which end on a Wilson line on $\gamma_1' \subset \mathfrak{B}_\text{phys}$.
    We can also terminate the $f_2$-flux tube emanating the Wilson line on a $\phi_1$-line on $V_p$, yielding a tube $\overline{W}_r$ which is anchored on both ends and topological if $r \in \mathbb{Z}$.
    Moreover we can join the lines on both sides to obtain the original lines $X_\ell$ which stretch between the two boundaries; we consider $\{X^L_m, X_\ell\}$ as the set of independent line operators.
    }
    \label{fig:X_split_Op}
\end{figure}

This construction applies for the 0-form $\mathbb{Z}_{k_A m_G}$ symmetry generators of individual universes with different fixed values of $f_0 = 2\pi m_G$.
Since this value jumps across the codimension-1 defects $V_p$, their insertion must also alter accordingly the set of charge and charged operators.
This is where the non-trivial topological sector on $V_p$ comes into play: by terminating tubes $W_\alpha$ and lines $X_n$ on the localized operators $v_q$ and $u_n$, respectively, we effectively obtain different operators on both sides of the $V_p$ insertion.
Let us now demonstrate this in detail.
For concreteness, we fix a choice of orientation by declaring that by inserting $V_p$ parallel to the boundary in the interior of the bulk, $f_0 = 2\pi m_G$ between $\mathfrak{B}_\text{sym}$ and $V_p$, which then jumps to $f_0 = 2\pi (m_G + p)$ between $V_p$ and $\mathfrak{B}_\text{phys}$, see Fig.~\ref{fig:X_split_Op}.

In this setting, we can have three qualitatively different gauge-invariant and topological tube operators built around a cylinder with $f_2$-flux, which are summarized in Fig.~\ref{fig:X_split_Op}.
First, we have
\begin{align}
    W^R_\alpha := \exp( i\alpha \oint_{\gamma_1} b_1+i q \int_{\mathcal{S}_2} f_2 - i q \oint_{\gamma_1'} \frac{a}{2\pi}) \, , \quad q = \alpha k_A (m_G + p) \in \mathbb{Z} \, ,
\end{align}
which is anchored on a Wilson line on the physical boundary and ends on a $b_1$-line in the bulk.
For now we assume that it does not cross $V_p$, so that $f_0 = 2\pi (m_G + p)$ along the entire tube; we will comment in a moment on what happens if we drag the $b_1$-line across the codimension-1 defect.
We can anticipate that these operators will generate the $\mathbb{Z}_{k_A (m_G + p)}$ 0-form symmetry in the universe with $f_0 = 2\pi (m_G +p)$ after collapsing the entire bulk.
An analogous tube can also be constructed on the other side of the defect (where $f_0 = 2\pi m_G$), with a free-floating $b_1$-line at the left end (which can be pushed all the way onto $\mathfrak{B}_\text{phys}$), but now anchored on a $\phi_1$-line on $V_p$,
\begin{align}
    W^L_\beta := \exp(i\beta \oint_{\gamma_1} b_1 + i b \int_{\mathcal{S}_2} f_2 - i b \oint_{\bar\gamma_1} \phi_1 ) \, , \quad b = \beta k_A m_G \in \mathbb{Z} \, .
\end{align}
Without the $V_p$ insertion, these tubes would instead end on Wilson-lines on the physical boundary and generate the $\mathbb{Z}_{k_A m_G}$ symmetry.
Finally, we can also have a third tube which is anchored on both ends: on a $\phi_1$-line on $V_p$ and a Wilson-line on $\mathfrak{B}_\text{phys}$,
\begin{align}
    \overline{W}_r := \exp( i r \oint_{\bar\gamma_1} \phi_1+i r \int_{\mathcal{S}_2} f_2 - i r \oint_{\gamma_1'} \frac{a}{2\pi}) \, , \quad r \in \mathbb{Z} \, .
\end{align}
We can now imagine dragging the $b_1$-line at the one end of $W_\alpha^R$ across the domain wall, which clearly is equivalent to joining a $W_\alpha^L$ and a $\overline{W}_q$ along a $\phi_1$-line.
However, for generic values of $m_G$ and $p$, the resulting tube $W_\alpha^L$ may not be topological because $\alpha k_A m_G$ need not be integer even if $\alpha k_A (m_G + p) \in \mathbb{Z}$.

To fully understand the 0-form symmetry, we must also talk about the lines operators $X_\ell$.
In the absence of $V_p$ in the bulk, $X_\ell$ stretches directly between the physical boundary and the symmetry boundary, giving rise to an operator charged under the 0-form symmetry. 
When $V_p$ is inserted in the middle, this $X_\ell$ line can now be split at the defect, as $X_\ell$ is allowed to terminate on the point operator $u_\ell = e^{i\ell \phi_0}$ localized on the defect. 
This gives lines on each side of the $V_p$ insertion (see Fig.~\ref{fig:X_split_Op}),
\begin{equation}
    \begin{aligned}
    X_m^L[\delta_1^L, y=\partial \delta_1^L ] &= \exp( im \int_{\delta_1^L} A_1 - im\phi_0(y) ) \, , \\ 
    X_n^R[\delta^R_1, z= \partial {\delta}^R_1 ] &= \exp( in\phi_0(z) + in \int_{\delta_1^R} A_1  ) \, ,
    \end{aligned}
\end{equation}
where we have suppressed the point operator $o_n$ localized on the physical boundary on which $X^R_n$ must terminate.
We proceed with $X_\ell$ (which stretch across the entire bulk) and $X_m^L$ as the independent set of line operators.

The linking between line and tube operators are straightforward to compute.
First, $X_\ell[\delta_1]$ links with both $W_\beta^L$ and $W_\alpha^R$ due to the canonical linking between the lines $\int_{\delta_1} A_1$ and $\int_{\gamma_1} b_1$ (see Fig.~\ref{fig:X_split_Op}); however, the phase differs due to the value of $f_0$ in the region where the $b_1$-lines sit:
\begin{equation}\label{eq:linking_XR-WR}
\begin{split}
    \langle X_\ell, W^R_\alpha \rangle & = \exp(2\pi i \alpha \ell) =  \exp[2\pi i \frac{\ell a}{k_A(m_G+p)} ] \qquad (a = \alpha k_A (m_G + p) \in \mathbb{Z})\, , \\
    \langle X_\ell, W^L_\beta \rangle & = \exp(2\pi i \beta \ell) =  \exp[2\pi i \frac{\ell b}{k_A m_G} ] \qquad (b = \beta k_A m_G \in \mathbb{Z})\, .
\end{split}
\end{equation}
On the left side of $V_p$, the linking between the operators $X_m^L[\delta_1]$, which extends from the symmetry boundary to the defect $V_p$, and $W^L_\alpha$ receives two contributions.
The first contribution, $e^{2\pi i m \alpha}$, comes from the linking of $b_1$-line of $W_\alpha^L$ with the $A_1$-line in $X^L_m$. 
The second factor, $e^{2\pi i \frac{m q}{p k_A}}$, comes from the linking of the $\phi_1$-line in $W_\alpha^L$ and the $\phi_0$-endpoint of $X^L_m$ which are the operators of the localized $\mathbb{Z}_{p k_A}$-BF-theory ${\cal A}$ \eqref{eq:worldvolume_Vp-action}.
This gives
\begin{equation}\label{eq:linking_XL-WL}
\begin{aligned}
    \langle X^L_m, W^L_\beta \rangle & = \exp[2\pi i \left(m \beta + \frac{mb}{pk_A}\right)] \qquad (b = \beta k_A m_G \in \mathbb{Z}) \\
    & = \exp[ 2\pi i \frac{m b}{k_A} \left(\frac{1}{m_G} + \frac{1}{p}  \right)] \\
    & = \exp[ 2\pi i m b \, \frac{\text{lcm}/m_G + \text{lcm}/p}{k_A \,\text{lcm}(m_G,p)} ] \, .
\end{aligned}
\end{equation}
It is clear from Fig~\ref{fig:X_split_Op} that $\langle X_m^L, W_\alpha^R \rangle = 1$.
Lastly, the tubes operators $\overline{W}_r$ link with $X^L_m$ but not $X_\ell$,
\begin{align}\label{eq:linking_Xs-What}
    \begin{split}
        \langle X^L_m, \overline{W}_r \rangle & = \exp(- 2\pi i \frac{m r}{pk_A} ) \, , \\
         \langle X_\ell, \overline{W}_r \rangle & = 1  \, .
    \end{split}
\end{align}

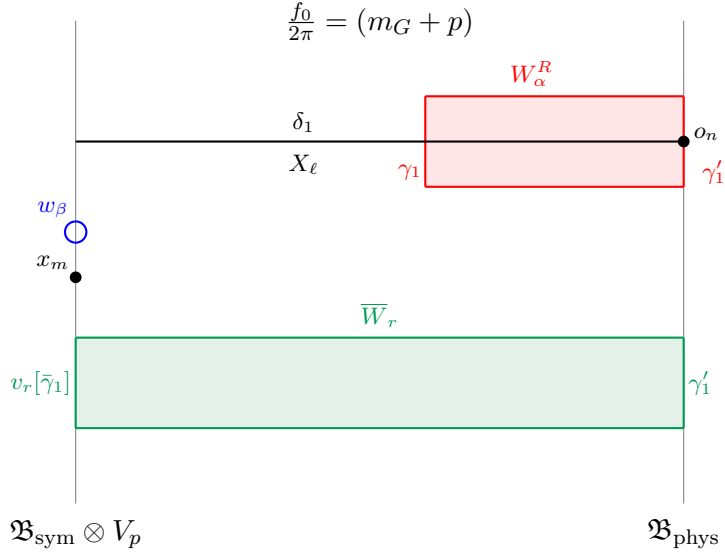
\begin{figure}[ht]
    \centering
    \begin{tikzpicture}[scale=2.0]
    \draw[gray] (-2,-.2) -- (-2,3);
    \draw[gray] (2,-.2) -- (2,3);

    \node at (0, 3) {{$\frac{f_0}{2\pi}=(m_G + p)$}};

    \node at (-2, -0.4) {$\mathfrak{B}_{\text{sym}} \otimes V_p$};
    \node at (2, -0.4) {$\mathfrak{B}_{\text{phys}}$};

    \draw[blue, thick] (-2, 1.6) circle (2pt);
    \node at (-2.15, 1.75) {\textcolor{blue}{\footnotesize{$w_\beta$}}};

    \filldraw (-2,1.3) circle (1pt);
    \node at (-2.15,1.4) {\footnotesize{$x_m$}};

    \fill[color=red!10] (0.3, 1.9) rectangle (2, 2.5);
    \draw[red, thick] (0.3,1.9) -- (0.3,2.5);
    \draw[red, thick] (2, 1.9) -- (2, 2.5); 
    \draw[red, thick] (0.3,1.9) -- (2,1.9); 
    \draw[red, thick] (0.3,2.5) -- (2,2.5);
    \node at (0.2, 2) {\textcolor{red}{\footnotesize{$\gamma_1$}}};
    \node at (2.2, 2) {\textcolor{red}{\footnotesize{$\gamma_1'$}}};
    \node at (1, 2.65) {\textcolor{red}{\footnotesize{$W^R_\alpha$}}};

    \draw[thick] (-2,2.2) -- (2,2.2);    
    \node at (-0.5, 2.05) {\footnotesize{$X_\ell$}};
    \node at (-.5, 2.33) {\footnotesize{$\delta_1$}};
    \filldraw (2,2.2) circle (1pt);
    \node at (2.15,2.25) {\footnotesize{$o_n$}};

    \fill[color=ForestGreen!10] (-2,0.3) rectangle (2,0.9);
    \draw[ForestGreen, thick] (-2,0.3) -- (2,0.3); 
    \draw[ForestGreen, thick] (-2,0.9) -- (2, 0.9);
    \draw[ForestGreen, thick] (-2,0.3) -- (-2,0.9);
    \draw[ForestGreen, thick] (2,0.3) -- (2,0.9);
    \node at (0, 1.05) {\color{ForestGreen}{\footnotesize{$\overline{W}_r$}}};
    \node at (- 2.23, 0.6) {\color{ForestGreen}{\footnotesize{$v_r[\bar\gamma_1]$}}};
    \node at (2.11, 0.6) {\color{ForestGreen}{\footnotesize{$\gamma_1'$}}};

\end{tikzpicture}
    \caption{The relevant operators after implementing the $(-1)$-form in the SymTFT, which is equivalent to pushing the $V_p$ insertion onto the symmetry boundary.
    The operators $W_\alpha^R$ and $X_\ell$ comprise the new $\mathbb{Z}_{k_A (m_G +p)}$ 0-form symmetry operators.
    Since $V_p$ contains a non-trivial TQFT, the resulting symmetry boundary is non-simple and contains topological point operators $x_m$ and line operators $w_\beta$ ($\beta k_A m_G = b \in \mathbb{Z})$.
    These two sectors are coupled by the presence of the tube operators $\overline{W}_r$ which are anchored on the enriched symmetry boundary and the physical boundary.
    We have the following linkings: $\langle x_m, w_\beta \rangle$ given by \eqref{eq:linking_XL-WL}, $\langle X_\ell, w_\beta \rangle = \exp( 2\pi i \frac{\ell b}{k_A m_G})$, $\langle x_m , \overline{W}_r \rangle = \exp(-2\pi i \frac{m r}{p k_A})$, and $\langle X_\ell , \overline{W}_r \rangle = 1$.
    }
    \label{fig:bulk_after_colliding_VpBsym}
\end{figure}

We finally have all the ingredients to describe the change in 0-form symmetry as a result of a $(-1)$-form symmetry transformation.
As we have seen for the invertible cases in Section \ref{sec:BF-SymTFT}, we should implement the $(-1)$-form symmetry transformation $f_0 \rightarrow f_0 + p$ by pushing the $V_p$ insertion into the symmetry boundary.
In the resulting bulk which has constant $f_0 = 2\pi (m_G + p)$, we have the expected $\mathbb{Z}_{k_A (m_G + p)}$ 0-form symmetry operators $X_\ell$ and $W_\alpha^R$ with the correct linking \eqref{eq:linking_XR-WR}.
However, this setup contains further operators localized on the enriched symmetry boundary $\mathfrak{B}_\text{sym} \otimes V_p$: there are topological point operators $x_m$ which are the result of squashing the $X^L_m$ lines, and topological line operators $w_\beta$ resulting from squashing the tubes $W_\beta^L$; they link with a $\mathbb{Z}_{k_A \text{lcm}(m_G, p)}$-valued phase \eqref{eq:linking_XL-WL}.
Since these operators are not attached to anything in the bulk, the resulting symmetry boundary $\mathfrak{B}_\text{sym} \otimes V_p$ is non-simple and in fact carries a 2d $\mathbb{Z}_{k_A \text{lcm}(m_G, p)}$-BF-theory.
Furthermore, this TQFT is not completely decoupled from the rest of the bulk and the physical boundary because of the $\overline{W}_r$ operators which extend into the bulk and link with $x_m$.
This setting is summarized in Fig.~\ref{fig:bulk_after_colliding_VpBsym}.

If we reduce the full interval, we end up the 2d theory with anomaly coefficient $k_A (m_G +p)$, but which is further stacked with the 2d $\mathbb{Z}_{k_A \text{lcm}(m_G,p)}$-BF-theory.
To recover the ``simple'' bulk-boundary system corresponding to the universe with $f_0 = m_G +p$, we can employ decomposition within the $\mathbb{Z}_{k_A \text{lcm}(m_G, p)}$-BF-theory.
That is, we can gauge the $(d-1)$-form symmetry generated by the newly obtained topological point operators $x_m$, i.e., projecting onto the invariant sector (possibly twisted by a discrete torsion parameter).
This procedure projects out the lines $w_\beta$ on the symmetry boundary since these link non-trivially with $x_m$, thus reducing it to a simple boundary.
Furthermore, it also projects out $\overline{W}_r$ with $r \not\equiv 0 \mod p k_A$.
Since the remaining $\overline{W}_r$ do not link with $X_\ell$ anyways, they are completely transparent, i.e., trivial operators in the theory after projection.
What then remains is the vanilla SymTFT for the operators of the $\mathbb{Z}_{k_A (m_G +p)}$ 0-form symmetry.

\section{Conclusions}
\label{conclusion}

In this work we analyzed $(-1)$-form symmetries and their quantum-dual $(d-1)$-form symmetries from the perspective of the SymTFT.
In this description, the parameter that the $(-1)$-form symmetry acts on, is identified with the Dirichlet boundary values of a locally constant scalar field $b_0$ and the associated topological point operator $U(x) = \exp(i \alpha b_0(x))$ in the SymTFT bulk, and is generated by $d$-dimensional topological domain walls $V(\Gamma_d)$ across which $b_0$ jumps.
By inserting these parallel to the symmetry boundary, we describe a $(-1)$-symmetry transformation by a spacetime-filling operator intersection after interval compactification of the SymTFT.
The fact that the dual theory with $(d-1)$-form symmetry --- obtained by swapping boundary conditions for $U$ and $V$ --- decomposes into universes which are related by $(-1)$-form symmetry transformations follows simply from a Fourier-like basis transformation on the SymTFT's Hilbert space that maps Dirichlet b.c.~for $V$ onto Dirichlet b.c.~for $U$ and vice versa.

We have demonstrated these aspects for group-like symmetries using the ``instanton restricted Yang--Mills'' model \cite{Tanizaki:2019rbk} (see also \cite{Yu:2024jtk, Najjar:2024vmm}), and have in particular generalized it by topologically gauging the full $U(1)^{(-1)}$ instanton symmetry of Yang--Mills to completely remove the sum over instanton sectors.
We have also presented how decomposition arises from gauging trivially acting $(d-2)$-form symmetries in the SymTFT.
We have contrasted the construction focusing on missing charges in the club sandwich approach \cite{Bhardwaj:2024qrf} with a symmetry-generator-centric perspective that has been employed in previous works on the subject \cite{Pantev:2005rh, Robbins:2020msp, Robbins:2021lry, Robbins:2021ibx, Pantev:2022kpl,Pantev:2022pbf, Robbins:2022wlr}.
In this context we have also highlighted an example where gauging a $(d-2)$-form symmetry \emph{does not} lead to decomposition, correcting a common misconception that the SymTFT helps address in a clean way.

Finally, motivated by the dimensional reduction of 4d $U(1)$ gauge theories with an ABJ-anomaly \cite{Nardoni:2024sos}, we have constructed a 3d SymTFT for a non-invertible $(-1)$-form symmetry in 2d.
The parameter $m_G$ associated with this symmetry encodes the 2d ABJ-anomaly coefficient, with which the discrete 0-form symmetry $\mathbb{Z}_{k_A m_G}$ changes.
To facilitate this change within the same SymTFT that contains also the charge(d) operators of the 0-form symmetry, the domain walls $V_p(\Gamma_2) \equiv V_p$ carry a non-trivial TQFT on their worldvolume with localized point and line operators on which the bulk surface and line operators associated with the 0-form symmetry can terminate.
By stacking the symmetry boundary with $V_p$, i.e., implementing the $(-1)$-form symmetry transformation $m_G \rightarrow m_G + p$, we find an enriched boundary with a 2d $\mathbb{Z}_{k_A \text{lcm} (m_G, p)}$-BF-theory which would be stacked on top of the 2d gauge theory with anomaly coefficient $k_A (m_G + p)$.
This is a direct consequence of the non-invertibility of the defects $V_p$.
To obtain ``just'' the 2d theory with modified anomaly coefficient, we can make use of the topological point operators associated with the BF-sector to construct a projector to an invariant sector, i.e., select one of the universes into which the BF-theory decomposes.

Clearly we have only scratched the surface of non-invertible $(-1)$-form symmetries.
In the example discussed here, we have focused on the defects $V_p$ whose construction was motivated by obtaining a gauge-invariant and genuine operator, which was sufficient for our purposes of demonstrating the change in 0-form symmetry.
However we have not made any attempt in clarifying the full categorical structure of the symmetry in this system.
Understanding this structure would be necessary to rigorously discuss the possibility of \emph{gauging} the $(-1)$-form symmetry, and study the resulting 2d theory which ought to decompose back into the various fixed gauge-flux sectors.
It would be also interesting to interpret this gauging in view of the 4d origin of this example where the sectors are simply a result of dynamical fluctuations of the 4d gauge field.

We also note that a $(-1)$-form symmetry transformation by means of a $V_p$ insertion bears strong resemblance to the club sandwich construction.
Indeed, the defect can be regarded as a special case of an interface which separates the bulk into two halves. 
However, in the club sandwich picture \cite{Bhardwaj:2023bbf} the interface separates a SymTFT $\mathfrak{Z}$ from another TQFT $\mathfrak{Z}' = \mathfrak{Z}/ {\cal A}$ with ${\cal A}$ a condensable algebra of $\mathfrak{Z}$, whereas here we have a (non-invertible, codimension one) defect within a single bulk theory.
Moreover, $V_p$ is also not a condensation defect \cite{Roumpedakis:2022aik, Choi:2022zal}, as there are point operators ($e^{i\beta f_0}$) which are charged under it.
If it were realizable via higher gauging, $V_p$ would instead be porous to point operators (like condensation defects obtained by 1-gauging a symmetry, there are line operators charged under the $V_p$'s, but we also have charged point operators).

Moreover, one may argue that in this example the non-invertibility associated with shifting the flux number $m_G \rightarrow m_G + p$ is ``mild'', in the sense that it effectively leads to finitely many copies of the theory with modified ABJ-anomaly due to decomposition on the non-simple symmetry boundary.
A possible conclusion, though unlikely, is that other non-invertible $(-1)$-form symmetries act in a similar way, i.e., the relevant physics being entirely captured by shifting the associated parameter.
To gain an insight into this, it may also be interesting to construct stringy models beyond those in \cite{Yu:2024jtk, Najjar:2024vmm, Najjar:2025htp}, to investigate possibly more exotic non-invertible $(-1)$-form symmetries in broader classes of models with a top-down construction.\footnote{One notable exception may be 3d ABJ(M) theories on M2-branes probing singular Calabi--Yau fourfolds, where the $(-1)$-form symmetry parameter is an 't Hooft anomaly coefficient for the 1-form symmetry \cite{vanBeest:2022fss, Yu:2024jtk}.
Promoting this coefficient to a dynamical field in the SymTFT bulk can potentially lead to a non-invertible codimension-1 defect.
We thank Xingyang Yu for sharing this observation.
}

As a final remark, recall that we derived the non-invertible $(-1)$-form symmetry with an simplifying assumption of $k_2$ (the 2-group ``structure constant'' of the 4d parent theory) being 0. 
If we allow $k_2$ to have non-zero values, it will introduce a mixed 't Hooft anomaly between the $(-1)$-form symmetry and the 0-form symmetry in each of the 2d universes. 
As the 0-form symmetry changes with the universe, the ``value'' of the 't Hooft anomaly has to change as well \cite{Nardoni:2024sos}. 
In that case, the $(-1)$-form symmetry has to account for the change of the 't Hooft anomaly along with the change in the 0-form symmetry, which certainly will require an even richer structure on the codimension-1 defects $V_p$ that remains to be explored.
We look forward to addressing these questions in future works.

\subsection*{Acknowledgments}
We are grateful to Dewi S.W.~Gould for initial collaboration on the project.
We further thank Ho Tat Lam, Alonso Perez-Lona, Eric Sharpe, Ethan Torres, Thomas Vandermeulen, and Xingyang Yu for helpful discussions and feedback.

\appendix

\section{\texorpdfstring{$\mathbb{Z}_N$}{Z-N} BF-theory and their boundary conditions}\label{app:symTFT_reviews}

Topological operators in a $(d+1)$-dimensional that satisfy the commutation relation \eqref{eq:commutation_BF-operators} can be modelled by a BF-theory.
For $G^{(p)} = \widehat{G}^{(d-p-2)} = \mathbb{Z}_N$, the corresponding action is\begin{align}\label{eq:BF-p-form}
    S^{(p)}_\text{bulk} = \frac{i N}{2\pi} \int_{{\cal W}_{d+1}} b_{d-p-1} \wedge d c_{p+1} \, ,
\end{align}
where $b$ and $c$ are $U(1)$-valued gauge fields.
The equations of motions constrain the topological operators
\begin{align}
    U(\Sigma_{d-p-1};k) = \exp(i k \textstyle\int_{\Sigma} b_{d-p-1}) \, , \quad V(\Gamma_{p+1};\ell) = \exp(i \ell \textstyle\int_{\Gamma} c_{p+1}) \, , \quad k,\ell \in \mathbb{Z}_N \, ,
\end{align}
to be $\mathbb{Z}_N \subset U(1)$-valued phases.
One can show that the linking property \eqref{eq:commutation_BF-operators} can be derived from canonically quantizing the fields $b$ and $c$ \cite{Witten:1998wy}.

By choosing the radial coordinate of ${\cal W}_{d+1} = {\cal M}_d \times [0,1]$ as the time coordinate in the quantization procedure, the boundary conditions at $r=0,1$ can be identified with states $|\mathfrak{B}_\text{sym} \rangle$ and $|\mathfrak{B}_\text{phys} \rangle$ of the Hilbert space.
The partition function of the $d$-dimensional QFT, obtained from interval reduction with such boundary conditions, is then given by
\begin{align}
    Z = \langle \mathfrak{B}_\text{phys} \, | \, \mathfrak{B}_\text{sym} \rangle \, .
\end{align}

Let us now discuss the boundary conditions.
For simplicity and our purposes, we shall only mention the canonical (``pure'') boundary condition, where all $U$ or all $V$ have the same type of boundary conditions.

To impose Dirichlet boundary conditions for $U$ or, equivalently, the field $b_{d-p-1}$, we introduce the boundary action
\begin{align}\label{eq:Dirichlet_b}
    S_\text{sym}[\lambda] = (-1)^{d-p-1} \, \frac{i N}{2\pi} \int_{\mathfrak{B}_\text{sym}} ( \hat{b}_{d-p-1} - d\lambda_{d-p-2} ) \wedge c_{p+1} \, ,
\end{align}
where $\lambda_{d-p-2}$ is a dynamical field on the boundary,\footnote{To maintain gauge invariance, $\lambda \rightarrow \lambda + \chi$ under a gauge transformation $b \rightarrow b + d \chi$.
Thus, $\lambda$ acts as a discrete St\"uckelberg field which spontaneously breaks the gauge symmetry of $b$ on the boundary.} and $\hat{b}_{d-p-1}$ is a \emph{fixed} background.
The equation of motion for $\lambda$, $N dc_{p+1} \equiv 0$, is the same as the bulk equation of motion for $b$.
However, the equation of motion for $c$ is modified by the boundary term, and we get the boundary condition
\begin{align}
    b_{d-p-1}|_{\mathfrak{B}_\text{sym}} =  \hat{b}_{d-p-1} - d \lambda_{d-p-2} \, .
\end{align}
That is, up to gauge transformations on the boundary $b$ is fixed to $\hat{b}$, which we can think of as the background field for the $\widehat{G}^{(d-p-2)}$ global symmetry.
In particular, the operators $U$ must end on ${\cal U}(\sigma_{d-p-2};k) = \exp(ik \int_\sigma \lambda)$, which are the charged objects of $\widehat{G}^{(d-p-2)}$.
As $c$ is not constrained on the boundary (other than the equations of motion inherited from the bulk), the topological operators $V$ can be parallel to $\mathfrak{B}_\text{sym}$, and generate $\widehat{G}^{(d-p-2)}$; this is typically referred to as Neumann boundary conditions for $V$ and $c$.

Dirichlet boundary conditions for $V$ and Neumann boundary conditions for $U$ is obtained in an analogous fashion, since the action \eqref{eq:BF-p-form} is symmetric under $b \leftrightarrow c$ upon integrating by parts and an unimportant overall sign depending on the degrees.
It is possible to also obtain Dirichlet boundary conditions for a subset of the $U$ and $V$ operators which give rise to mutually local operators ${\cal U}$ and ${\cal V}$ on the boundary; for the sake of our discussion we will only not concern ourselves with such mixed boundary conditions.

Assuming that we can have both of these two canonical boundary conditions with a fixed choice of physical boundary conditions $\mathfrak{B}_\text{phys}$, we have two absolute theories with different global symmetries and their background fields:
\begin{equation}
    \begin{aligned}
        & \mathfrak{B}^{(\hat{c})}_\text{sym} \, : \quad && c_{p+1}|_{\mathfrak{B}_\text{sym}} \equiv \hat{c}_{p+1} \quad && \Rightarrow \, G^{(p)} \text{ global symmetry,} \\
        & \mathfrak{B}^{(\hat{b})}_\text{sym} \, : \quad && b_{d-p-1}|_{\mathfrak{B}_\text{sym}} \equiv \hat{b}_{d-p-1} \quad && \Rightarrow \, \widehat{G}^{(d-p-2)} \text{ global symmetry.}
    \end{aligned}
\end{equation}

\bibliographystyle{JHEP}
\bibliography{refs.bib}
\end{document}